\documentclass[12pt,preprint]{aastex}
\usepackage{epsfig}
\begin{document}

\title{Candidate Type II Quasars from the Sloan Digital Sky Survey: \\
I. Selection and Optical Properties of a Sample at $0.3<Z<0.83$}

\author{
Nadia L. Zakamska\altaffilmark{1}, 
Michael A. Strauss\altaffilmark{1}, 
Julian H. Krolik\altaffilmark{2},  
Matthew J. Collinge\altaffilmark{1},\\
Patrick B. Hall\altaffilmark{1}, 
Lei Hao\altaffilmark{1}, 
Timothy M. Heckman\altaffilmark{2},
\v{Z}eljko Ivezi\'c\altaffilmark{1},\\ 
Gordon T. Richards\altaffilmark{1}, 
David J. Schlegel\altaffilmark{1},
Donald P. Schneider\altaffilmark{3},
Iskra Strateva\altaffilmark{1},\\
Daniel E. Vanden Berk\altaffilmark{4},
Scott F. Anderson\altaffilmark{5},
Jon Brinkmann\altaffilmark{6}}
\altaffiltext{1}{Princeton University Observatory, Princeton, New Jersey 08544}
\altaffiltext{2}{Department of Physics and Astronomy, Johns Hopkins University, 3400 North Charles Street, Baltimore, MD 21218-2686}
\altaffiltext{3}{Department of Astronomy and Astrophysics, 525 Davey Laboratory, Pennsylvania State University, University Park, PA 16802}
\altaffiltext{4}{Department of Physics and Astronomy, University of Pittsburgh, 3941 O'Hara st., Pittsburgh, PA 15260} 
\altaffiltext{5}{Department of Astronomy, University of Washington, Box 351580, Seattle, WA 98195}
\altaffiltext{6}{Apache Point Observatory, P.O. Box 59, Sunspot, NM 88349}

\begin{abstract}

Type II quasars are the long-sought luminous analogs of type II (narrow emission line) Seyfert galaxies, suggested by unification models of active galactic nuclei (AGN) and postulated to account for an appreciable fraction of the cosmic hard X-ray background. We present a sample of 291 type II AGN at redshifts $0.3<Z<0.83$ from the spectroscopic data of the Sloan Digital Sky Survey. These objects have narrow (FWHM$<2000$ km s$^{-1}$), high equivalent width emission lines with high-ionization line ratios. We describe the selection procedure and discuss the optical properties of the sample. About 50\% of the objects have [OIII] $\lambda_{\rm air}$5007 line luminosities in the range $3\times 10^8-10^{10}L_{\odot}$, comparable to those of luminous ($-27<M_B<-23$) quasars; this, along with other evidence, suggests that the objects in the luminous subsample are type II quasars. 
\end{abstract}


\section{Introduction}
\label{introduction}

The current understanding of the collective properties of active galactic nuclei (AGN) can be summarized by so-called unification models (e.g., \citealt{anto93, urry95}) in which the observed properties of active galaxies are governed primarily by orientation and intrinsic luminosity. For low luminosity AGN (Seyfert galaxies) a wide range of properties can be explained by differences of viewing angle to a nucleus surrounded by dusty obscuring material. If the viewing angle is such that the innermost region is not shielded from the observer, the optical spectrum is characterized by strong blue continuum and broad emission lines. However, if the line of sight happens to pass through a sufficient amount of obscuring material, the nuclear spectrum is dominated by narrower emission lines emitted by gas farther from the central engine. In this picture, the difference in line widths is naturally related to the depth of the potential well that the observer probes at different distances from the central object. More specifically, obscured (type II) Seyfert galaxies manifest themselves as narrow high-ionization emission line objects and outnumber unobscured broad line (type I) Seyfert galaxies in the local universe \citep{oste88, salz89, huch92, keel94, ho97, krol99}.  

Unification models predict that AGN output a significant fraction of their bolometric luminosity in the far-infrared because part of the UV and optical continuum is thermally reprocessed by circumnuclear dust. In obscured (type II) AGN the strong optical continuum is hidden from the observer, so the observed far-IR-to-optical ratios are expected to be large. These objects are also expected to have hard X-ray spectra because the gas along the line of sight absorbs soft X-rays and permits only the hard X-rays from the nucleus to escape directly. 

If the same unification models apply to higher-luminosity AGN (quasars), there should exist high-luminosity obscured AGN (type II quasars) which would be observable up to high redshifts. Obscured AGN have been postulated to account for the cosmic hard X-ray background \citep{mada94, zdzi95}, and type II quasars might conveniently account for a large fraction of it. A number of different methods have been used hitherto to discover type II quasars, but only a handful have been found. 

In the optical, type II quasar candidates have been selected as objects with narrow permitted emission lines and high-ionization line ratios. \citet{djor01} selected several candidates at $Z=0.31-0.36$ as outliers from the standard color-color loci of stars, galaxies and quasars from the Digital Palomar Observatory Sky Survey. 

\citet{klei88} selected one type II quasar candidate at $Z=0.442$ as a powerful IR source. This object was later confirmed to possess properties expected from a type II quasar in the optical and in the X-rays \citep{fran00, iwas01}. 

Radio-loud type II AGN (narrow-line radio galaxies) are readily detectable in radio surveys and have been known for decades (see \citealt{mcca93} for a review) out to very high redshifts \citep{vanb99}. A typical high-redshift radio galaxy requires a hidden quasar with a luminosity $10^{46}-10^{47}$ erg s$^{-1}$ in order to explain its powerful optical emission lines \citep{mcca93}. 

Several type II quasar candidates have been selected as objects with unusually hard X-ray spectra (\citealt{daws01, norm01, ster02, dell03} at redshifts 2.01, 3.7, 3.29, and 0.40, respectively). Multi-wavelength observations of these objects support their identifications as type II quasars. In particular, in the optical they show high equivalent width, narrow emission lines with high-ionization line ratios. However, in many other cases follow-up optical observations of promising X-ray selected candidates revealed broad components in their permitted emission lines \citep{shan96, halp99, akiy02}. Overall, no more than 20\% of hard X-ray sources in deep surveys have high-ionization narrow emission line optical counterparts (e.g., \citealt{barg01, gand02}; or \citealt{matt02} for an overview). In general, X-ray and optical selection techniques, while complementing each other, do not necessarily reveal the same populations of objects. X-ray selection favors high-redshift objects, for which their hard X-ray spectra are redshifted into the observable range, so candidates found by X-ray surveys can be very faint optically (down to $m_R=25$). Obtaining optical spectra of these candidates is thus a challenging task. On the other hand, very Compton-thick type II AGN can be optically bright while being quite faint in X-rays \citep{bass99}. 

The Sloan Digital Sky Survey (SDSS; \citealt{york00}) makes it possible to find a large number of type II quasar candidates. In this paper we present a large sample (291 objects) of high-ionization narrow emission line AGN with $0.3<Z<0.83$ from the SDSS spectroscopic data. The sample includes Seyfert II galaxies as well as objects luminous enough to be called type II quasar candidates. We briefly summarize relevant information on the SDSS imaging, targeting for spectroscopy and processing of spectroscopic data in Section \ref{data}. We describe our selection procedure for type II AGN in Section \ref{selection}, and present the sample in Section \ref{properties}, where we discuss the collective and individual observed properties of the objects from our sample. In Section \ref{composite} we present the composite spectrum of type II AGN and we estimate their intrinsic luminosities in Section \ref{discussion}. We summarize our results in Section \ref{conclusions}.

The SDSS spectroscopic survey offers a unique opportunity to study large samples of AGN with well-defined selection criteria. Spectroscopic data for about 16,000 quasars, and many more lower-luminosity AGN, are now publicly available \citep{abaz03, schn03}. The high quality of the spectroscopic data and the size of the survey allow unprecedented study of the properties of AGN. In particular, in our research we make extensive use of results from the study of nearby type II AGN and their host galaxies by \citet{kauf03} and from the study of emission-line galaxies and nearby AGN by \citet{hao03}. In addition to selecting large homogeneous samples of extragalactic objects, the survey is optimized for the discovery of rare and unusual objects. The combination of the size of the survey and the `serendipity' search program resulted in finding numerous unusual AGN, such as extreme BAL quasars \citep{hall02}, extremely red quasars \citep{rich03}, BL Lac objects \citep{ande03}, and AGN with double-peaked broad emission lines \citep{stra03}, and makes it possible to find a significant number of type II quasars, as we describe in this paper.

Luminosities are calculated in a $h=0.7$, $\Omega_m=0.3$, $\Omega_{\Lambda}=0.7$ cosmology (also referred to in this paper as the $\Lambda$-cosmology). We use IAU designations and J2000 coordinates for identifying objects (e.g., SDSS J102746.03$+$003205.0). To avoid confusion, we use capital $Z$ for redshifts and lower-case $z$ for the $z$-band magnitudes. All wavelengths of the emission lines in the text are vacuum wavelengths taken from the Atomic Line List\footnote{The Atomic Line List is hosted by the Department of Physics and Astronomy at the University of Kentucky (http://www.pa.uky.edu/$\sim$peter/atomic)}; the conventionally used air wavelengths are listed in Table 2. All equivalent widths (EWs) and all full widths at half maximum (FWHM) are rest-frame values. 

\section{SDSS Observations and Data Processing}
\label{data} 

The Sloan Digital Sky Survey (SDSS; \citealt{york00}) uses a drift-scanning imaging camera \citep{gunn98} and a 640 fiber double spectrograph on a dedicated 2.5m telescope. It is an ongoing survey to image 10,000 sq. deg. of the sky in the SDSS $ugriz$ AB magnitude system \citep{fuku96, lupt99, stou02} and to obtain spectra for $\sim 10^6$ galaxies and $\sim 10^5$ quasars. The astrometric calibration is good to better than $0\farcs 1$ rms per coordinate \citep{pier03}, and the photometric calibration is accurate to 3\% or better \citep{hogg01, smit02}. We use Petrosian magnitudes for resolved sources and PSF magnitudes for unresolved sources as measures of flux; asterisks on magnitudes indicate that preliminary photometry is used (see description in \citealt{stou02, abaz03}).

We now proceed to describe targeting for spectroscopy. About 66\% of the main spectroscopic survey is the galaxy sample. Resolved sources are targeted as galaxies, down to a limiting magnitude of $r=17.77$ \citep{stra02}. Luminous red galaxies (LRGs) constitute about 9\% of the spectroscopic survey and are targeted based on their distinctive colors down to a limiting magnitude of $r=19.5$ \citep{eise01}. Another 13\% of the survey are quasar candidates targeted based on their colors down to $i=19.1$, plus additional high-redshift candidates ($Z>3$) which are targeted to $i=20.2$ \citep{rich02}. Objects in the few untargeted regions of color space outside the main stellar locus, plus objects with radio emission detected by the FIRST survey \citep{beck95, ivez02}, plus objects with X-ray emission detected by the ROSAT survey \citep{voge99, ande03} can also be selected to $i=20.5$ as `serendipity' targets \citep{stou02} which constitute about 4\% of the spectroscopic survey. Fibers are allocated according to a tiling algorithm \citep{blan03} with the galaxy sample and the quasar sample being the top priorities. The remaining 8\% of fibers serve for calibration purposes. 

A small number of plates are devoted to miscellaneous projects that are not part of the main spectroscopic survey; for such plates, target selection algorithms different from those discussed above were applied. For some plates, hereafter designated ``Special plates'', resolved objects with $17.7<r<19.5$ were targeted, with a preference to objects bluer than normal galaxies (on average, about 1 mag bluer in $u^*-g^*$ than the main galaxy sample). Most objects on the Special plates are emission-line galaxies. A few plates are dedicated to the spectroscopic follow-up of the Deep Southern Equatorial Scan (DSES; a project for repeat imaging of certain fields in the southern portion of the SDSS area). For these plates, hereafter designated ``DSES plates'', the Quasar, High-redshift quasar and Serendipity selection algorithms select objects somewhat fainter (up to 1 mag) than in the main spectroscopic sample.  

SDSS spectra are obtained using plates holding 640 fibers, each of which subtends 3$\arcsec$ on the sky; the spectra cover $3800-9200\mbox{\AA}$ with resolution of $1800<R<2100$ and sampling of $\simeq 2.4$ pixels per resolution element. The relative and absolute spectrophotometric calibration is good to $\sim 10-20\%$. Some objects have multiple spectra, either because they lie on multiple plates (which is done for quality assurance purposes) or because a plate was observed more than once. The spectra presented in this paper are co-additions of all available spectra using inverse variance weighting at each pixel. Spectral flux errors per pixel in most cases are about $1\times 10^{-17}$ erg s$^{-1}$ cm$^{-2}$ \AA$^{-1}$.

Redshifts are automatically assigned by the SDSS spectral classification algorithm, which is based on $\chi^2$ fitting of templates to each spectrum (Schlegel et al., in preparation). Averaged over the whole survey, about 99\% of redshift determinations are correct, as confirmed by visual inspection. 

\section{Spectroscopic Selection Criteria}
\label{selection}

In this Section we describe the selection procedure aimed at finding objects with optical properties expected of obscured AGN in the SDSS spectroscopic data. We search for objects with narrow emission lines without underlying broad components and with line ratios characteristic of non-stellar ionizing radiation. We discuss how we distinguish them from star-forming galaxies, in particular low-metallicity emission line systems, and narrow-line Seyfert 1 (NLSy1) galaxies (e.g., \citealt{will02}).

\subsection{Initial criteria}

In this paper we discuss objects with redshifts in the range $0.3<Z<0.83$. Detailed studies of local ($Z<0.33$, $r<17.7$ mag) AGN from the SDSS are presented by \citet{kauf03} and \citet{hao03}.  We only considered objects with $Z>0.3$ to disfavor selection of low-luminosity objects. We restricted the analysis in this paper to redshifts $Z<0.83$ in order to have the [OIII] 5008 line (the strongest expected emission line) present in all spectra, to be able to study the emission line properties of the objects statistically. (We use vacuum wavelengths throughout this paper, but air wavelenths are listed in Table 2 for convenience.)

We required signal-to-noise ratios (S/N) $\geq 7.5$, where by ``signal'' here we mean the flux density in the 7th brightest pixel over the whole spectroscopic range (about 3840 pixels) and by ``noise'' we mean the median estimated flux error per pixel over all pixels. This unconventional criterion allows us to keep objects with weak continua but strong narrow emission lines, while rejecting low S/N continuum sources. For illustration, an emission line with FWHM$=600$ km s$^{-1}$ occupies about 7 pixels of the spectroscopic range. We avoid using the brightest pixels because they can be strongly affected by sky subtraction and other instrumental problems. 

In order to select emission line objects, we required the rest-frame equivalent width of the [OIII] 5008 line to be $\geq 4\mbox{\AA}$.

\subsection{Narrow permitted emission lines} 
\label{line_width}

The division in line width between narrow-line and broad-line AGN is not well determined in the literature; moreover, the width criteria are applied to different permitted lines depending on which are available in the spectrum. The specific FWHM criteria range from 1000 km s$^{-1}$ (e.g., \citealt{weed77}) to 2000 km s$^{-1}$ (e.g., \citealt{stei02}). \citet{hao03} provided a formal separation criterion by showing that the width distribution of the H$\alpha$ line is strongly bimodal among AGN, with a significant dip at FWHM(H$\alpha$)$=1200$ km s$^{-1}$. 

Lines can be broadened by splitting, outflows or other effects, as discussed in Section \ref{unusual} below. To retain such unusual objects, we adopted the more conservative criterion FWHM(H$\beta$)$<2000$ km s$^{-1}$. However, this criterion by itself does not reject some genuinely unobscured objects, in particular, NLSy1 galaxies (e.g., \citealt{will02}). Our high-ionization selection criteria are designed to disfavor selection of these objects based on the emission line ratios, as described in Section \ref{high-ion}. Below we show that the results of our selection are not very sensitive to the exact value used for the line width cut. 

At the redshifts of our sample ($Z>0.3$) the projected aperture of the fiber (3$\arcsec$) is greater than 13.4 kpc, and thus the spectra include most of the light from the host galaxy. The emission lines coming from the nuclei can be significantly contaminated by stellar absorption lines from the host galaxy, so we subtract the host galaxy contribution from each spectrum before measuring the line parameters following \citet{hao03}. To perform this subtraction, we fit each spectrum to a linear combination of several templates and minimize the residuals to find the best-fit combination. We use the following templates: (a) the first seven Principal Component Analysis (PCA) components derived from a large sample of high S/N, pure absorption line SDSS galaxies; (b) an A-type star template to account for recent star formation in the host galaxy and (c) a power-law component proportional to $\lambda^{-1.5}$. We have to use the A star template explicitly because the pure absorption-line galaxies used in the PCA contain only old stellar populations (galaxies with a lot of young stars are likely to show emission lines from associated HII regions). The power-law component is introduced to represent any continuum contribution from the nucleus and, possibly, very young O and B stars. The fitting procedure is applicable to rest wavelengths $\lambda>3600\mbox{\AA}$ for which there exist reliable galaxy templates. In the final fit, the stellar contribution (stars of the spectral type A and later) is the linear combination of templates (a) and (b). 

We subtract the stellar contribution from the original spectrum and then proceed with characterizing the emission lines. We fit all emission lines with Gaussian profiles. A constant fit to the continuum in the vicinity of the line is used. The best-fit height and the width of the profiles, the wavelength of the peak of the flux, as well as the continuum level are obtained by minimizing $\chi^2$. 

Our goal was to reject all objects with broad components in their emission lines. Due to the spectroscopic coverage, the strong permitted lines that are available for study are H$\alpha$ and H$\beta$ for $0.3<Z<0.4$ and H$\beta$ and MgII 2800 for $0.4<Z<0.83$. Many objects show weak broad components in H$\beta$ and/or H$\alpha$, in addition to much stronger narrow components; automated procedures attempting to detect underlying weak broad components were found to be inefficient because the narrow lines often show deviations from the Gaussian shape (e.g., asymmetry or splitting). We fitted the H$\alpha+$[NII] 6550, 6585 and H$\beta+$[OIII] 4960, 5008 line complexes with four Gaussians. The [NII] 6550, 6585 lines are fitted with Gaussian profiles of the same width and with the height ratio of 1/3 as required by the energy level structure of the [NII] ion \citep{oste89}. [OIII] 4960, 5008 are fitted in the same way. We fitted Balmer lines with two Gaussian components and excluded all objects with strong broad components, i.e. whose broad component had a FWHM of at least 2000 km s$^{-1}$ and the same amplitude (peak $f_{\lambda}$) as the narrow component or larger. We then visually inspected all other spectra and rejected all objects where the broad component could be detected above the noise level. 

MgII 2800 requires a different fitting procedure. Because the narrow component of MgII 2800 is generally quite weak compared to the broad component (if the latter is present), we fit MgII 2800 with only a single component and reject all objects with FWHM$>$2000 km s$^{-1}$ if the amplitude of the line is detected above the noise level.

However, a broad component with amplitude comparable with the noise (i.e., about $10^{-17}$ erg s$^{-1}$ cm$^{-2}$ \AA$^{-1}$) cannot be detected by any of our procedures; it is therefore possible that deeper spectroscopy could reveal broad components in some of the objects in our sample. Whenever available, we combined (using error-weighting) multiple observations of the same object to obtain a higher S/N spectrum, and rejected objects which have a broad component in Balmer lines or in MgII 2800 in the co-added spectrum. This procedure reveals weak broad components in about 6\% of objects with multiple spectra. To determine whether the sample remains significantly contaminated by objects with weak broad components, we constructed a composite spectrum of the entire sample (see Section \ref{composite}). There remains a possibility that for $Z>0.40$ there is a broad component in H$\alpha$ which is redshifted out of the spectral range, while there is none at an observable level in H$\beta$. However, at these redshifts MgII 2800 is present in the spectra and can be used as an additional check. Whether some of objects at $Z>0.4$ have broad components in H$\alpha$ (while only having a narrow MgII 2800) can only be resolved with near-IR spectroscopy.

\subsection{High ionization emission lines} 
\label{high-ion}

The idea to use emission line ratios to distinguish between ionization mechanisms goes back to \citet{bald81} (and references therein), with high-ionization line ratios being indicative of photoionization by the underlying power-law AGN continuum. To distinguish type II AGN from other narrow emission line objects (star-forming galaxies and NLSy1 galaxies) we used diagnostic diagrams in the form suggested by \citet{kewl01}. The SDSS spectroscopic database provides a large number of emission-line galaxies which enables detailed statistical studies of the properties of the objects in different regions of the diagnostic diagrams \citep{hao03, kauf03}. We measure the relative fluxes of several emission lines (H$\beta$, [OIII] 5008, H$\alpha$, [NII] 6585 and [SII] 6718, 6733) and then apply the AGN diagnostic, which in analytical form is 
\begin{eqnarray}
\log\left(\frac{\rm [OIII]5008}{\rm H\beta}\right)>\frac{0.61}{\log({\rm [NII]/H\alpha})-0.47}+1.19;\label{diag1}\\
\log\left(\frac{\rm [OIII]5008}{\rm H\beta}\right)>\frac{0.72}{\log({\rm [SII]/H\alpha})-0.32}+1.30.\label{diag2}
\end{eqnarray}
\citet{kewl01} and \citet{hao03} use an additional criterion involving the [OI] 6302 line; we do not employ it in our selection because this line is weak, and for our redshift coverage it lies in the red part of the spectrum where sky subtraction is particularly difficult, making this line an unreliable measure.

We use criteria (\ref{diag1})-(\ref{diag2}) for objects at redshifts $0.3<Z<0.4$. For $Z>0.37$, the [SII] 6718, 6733 lines are redshifted out of the observed range so that criterion (\ref{diag2}) can no longer be applied. Finally, at $Z>0.40$ H$\alpha$ moves out of the spectroscopic range, so criterion (\ref{diag1}) cannot be used either. For $Z>0.40$ we adopt the following criteria:
\begin{enumerate}
\item The [OIII] 5008/H$\beta$ ratio requirement in reduced form:
\begin{equation}
\log\left(\frac{\rm [OIII]5008}{\rm H\beta}\right)>0.3.
\label{reduced}
\end{equation}
As can be seen from the diagnostic diagrams of \citet{kewl01} and \citet{hao03}, AGN satisfy this criterion, but the majority of objects that meet this criterion at low redshift are low-metallicity emission-line galaxies. We restricted our selection to $Z>0.3$, so we disfavor selection of star-forming galaxies relative to AGN due to the intrinsically lower luminosities of the former. Nevertheless, criterion (\ref{reduced}) alone is not sufficient to distinguish type II AGN from star-forming galaxies.
\item In addition, a sign of AGN activity should be present: either (a) the high-ionization lines [NeV] 3347, 3427 should be detected, or (b) FWHM([OIII])$>400$ km s$^{-1}$. The [NeIII] 3870 line can be prominent in the spectra of star-forming regions \citep{rola97}, but the higher ionization [NeV] 3347, 3427 lines are unambiguously characteristic of AGN activity. When low S/N does not allow detection of [NeV] lines, we rely on the correlation between line width and ionization ratios established by \citet{hao03}. This correlation implies that the majority of AGN have FWHM([OIII])$>400$ km s$^{-1}$, while star-forming galaxies have FWHM([OIII])$<400$ km s$^{-1}$. Low-metallicity emission line systems can be a source of confusion due to their large [OIII] 5008/H$\beta$ ratios \citep{ugry03, knia03} but they have FWHM([OIII])$<200$ km s$^{-1}$.
\end{enumerate}

Although originally the criteria of \citet{kewl01} were only used to distinguish AGN from star-forming galaxies, we found the high-ionization criteria very useful in rejecting NLSy1 galaxies, which are characterized by narrow H$\beta$ (FWHM$<2000$ km s$^{-1}$), weak [OIII] 4960, 5008 and strong FeII lines (e.g., \citealt{will02} and references therein). In fact, one observational definition of NLSy1 includes the criterion [OIII]5008/H$\beta<3$ \citep{halp98}. Reassuringly, not a single NLSy1 object from the SDSS Early Data Release catalogued by \citet{will02} met the high-ionization selection criteria presented here. We also do not see any sign of the FeII emission blend around $4750\mbox{\AA}$ in any of the type II AGN candidates we selected. Because the high-ionization criteria efficiently reject NLSy1, the results of our selection are remarkably insensitive to the line width cut that we set at FWHM(H$\beta$)$<2000$ km s$^{-1}$, with only a few objects having 1000$<$FWHM(H$\beta$)$<2000$ km s$^{-1}$. We further address possible contamination of our sample by NLSy1 in Section \ref{comp_contaminants}.

Finally, although we do not rely on the [NeIII] 3870 line in our selection, we find that objects that we select generally fall in the AGN region of the diagnostic diagrams involving [NeIII] 3870 from \citet{rola97} (the diagrams of [NeIII]/H$\beta$ vs [OII]/H$\beta$ and [OIII]/H$\beta$).

The selection algorithm was run over 600 plates, observed as of 22 August 2002, of which 291 plates are available in the Data Release 1 \citep{abaz03}. These plates contain 384,000 spectra; 40,379 objects are automatically classified as having redshifts in the range $0.3<Z<0.83$. Of these, we selected 291 type II AGN candidates according to the selection criteria described here.

\section{Optical properties of the sample}
\label{properties}

The sample of type II AGN at redshifts $0.3<Z<0.83$ is presented in Table 1, with their photometric and spectroscopic properties (the Table appears in full in the electronic version of the Journal only). The first column is an identifier in the IAU designations which includes the J2000 coordinates. The column ``DR1'' indicates whether the spectrum of the object is available in the SDSS Data Release 1 \citep{abaz03}. The redshifts given in the table are based on the [OII] 3728 emission line. This line is a close doublet that we fit with one Gaussian profile assuming that the centroid of the line in the rest frame is at $\lambda=3728.4\mbox{\AA}$. The resulting redshift error is $\sigma_Z\leq 0.001$.

The targeting information is in the $5+2$ digit target code that shows how the object was targeted for spectroscopy. The first five digits indicate whether the object was targeted on one of the main spectroscopic plates by the Galaxy, LRG, Quasar, High-redshift quasar or Serendipity algorithms, in that order. A digit of 0 means the object was not targeted by this method, while 1 means targeted. The algorithms have different selection criteria and limiting magnitudes, as described in Section \ref{data}. The last two digits show whether the object was targeted on one of the DSES or one of the Special plates. 

The next five columns give PSF magnitudes for the objects in the sample. An object should be considered properly detected in a particular band if the magnitude error in this band is $<0.2$ mag. Emission from the nucleus, central parts of the host galaxy, and possibly circumnuclear starburst contribute to the PSF magnitudes of AGN, but using these magnitudes minimizes contamination by the host galaxy.

The next two columns give rest-frame equivalent widths of the brightest emission lines, [OII] 3728 and [OIII] 5008. The following two columns give the logarithms of luminosities for these emission lines in solar units as calculated after fitting the lines with Gaussians. The column ``FIRST'' gives the integrated flux density in mJy at 20 cm \citep{ivez02} from the FIRST survey for objects with FIRST matches within 3$\arcsec$. If the field of the object has not been observed in the FIRST survey, this column lists `n/a'. In the column ``comments'', the flag `asym' is set for the objects that show unusual (asymmetric) [OIII] 5008 emission lines. These objects are discussed in Section \ref{unusual}.  

\begin{figure}
\epsscale{0.9}
\plotone{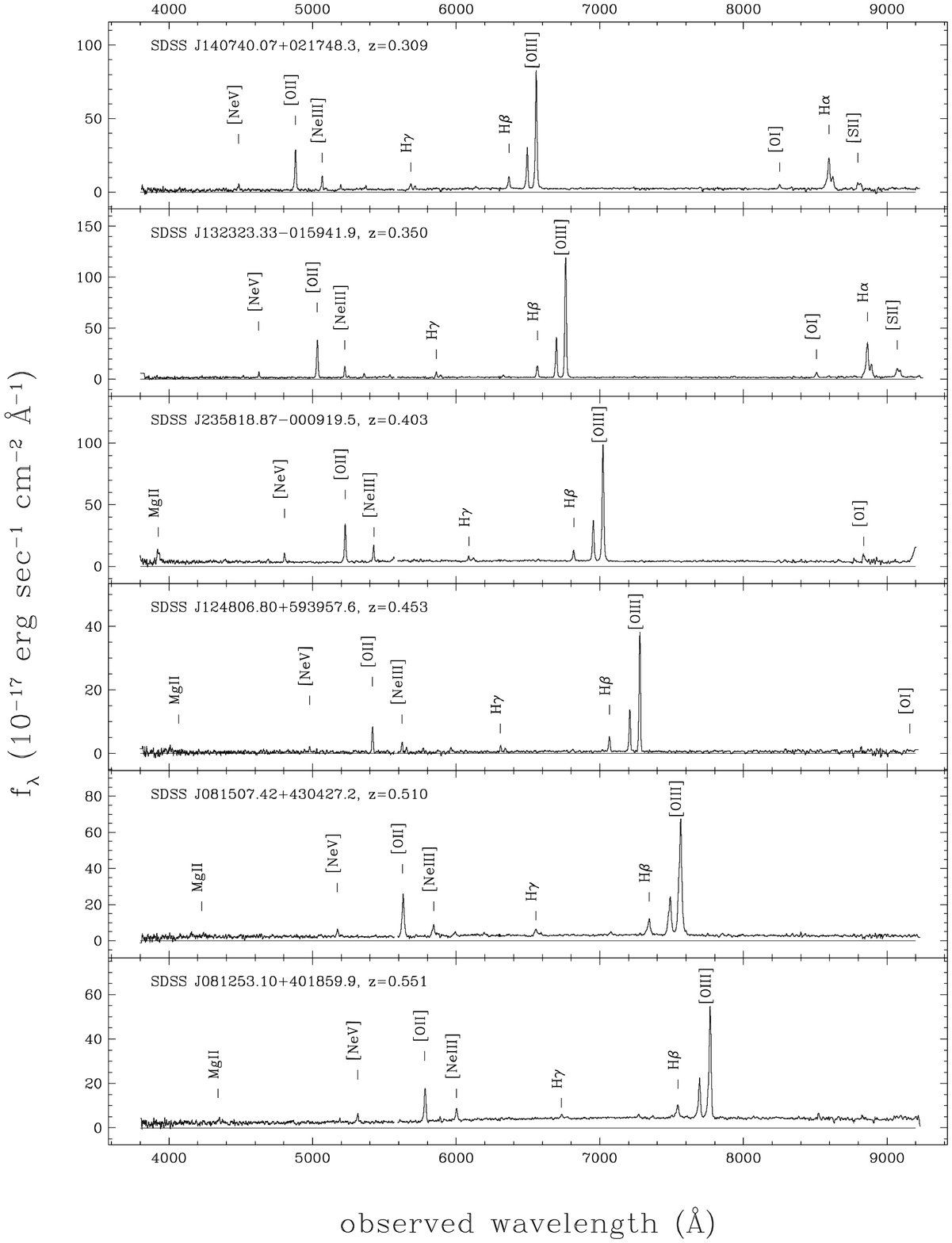}
\figcaption{Example spectra of type II quasar candidates, $0.3<Z<0.6$, smoothed by 5 pixels. The thin line is at the zero flux density level. For this Figure, objects with high line luminosities were chosen ($L$[OIII]$>3\times 10^8L_{\odot}$, see Section \ref{3108}).\label{example1}}
\end{figure}

\begin{figure}
\epsscale{0.9}
\plotone{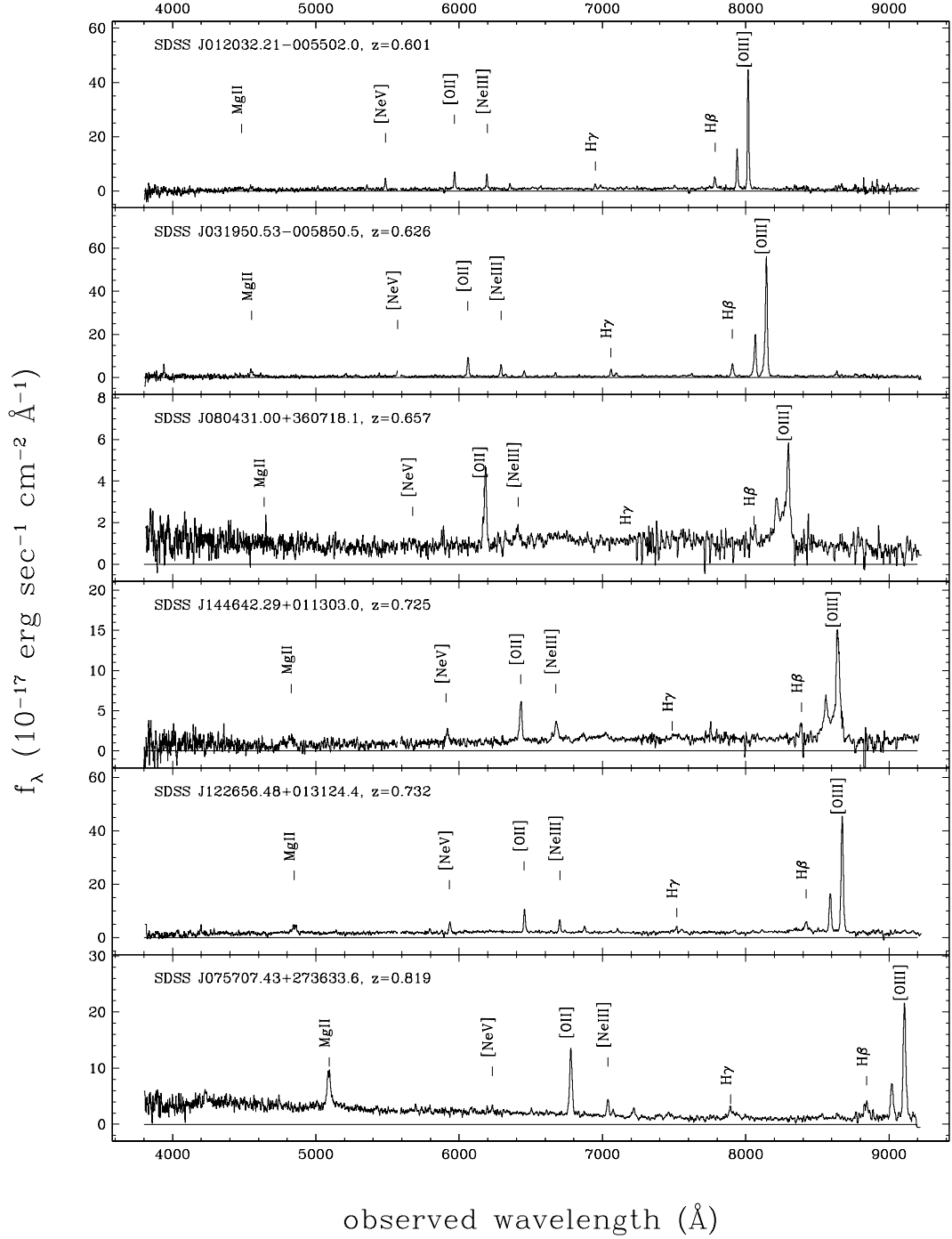}
\figcaption{Same as Figure \ref{example1}, for redshifts $0.6<Z<0.83$.\label{example2}}
\end{figure}

Figures \ref{example1} and \ref{example2} show example spectra of type II quasar candidates in order of increasing redshift; for these figures we chose objects with high [OIII] 5008 line luminosities ($L{\rm [OIII]}>3\times10^8 L_{\odot}$, see Section \ref{3108}). Major emission lines used for classification are labeled. Noteworthy features of the spectra are the narrow high-ionization emission lines (e.g., [NeIII] 3870 and [NeV] 3347, 3427) and high [OIII]/H$\beta$ ratios; the continuum is virtually absent, although in some cases the presence of the host galaxy can be traced when CaII 3935, 3970 absorption lines are detected. 

\begin{figure}
\epsscale{1.0}
\plotone{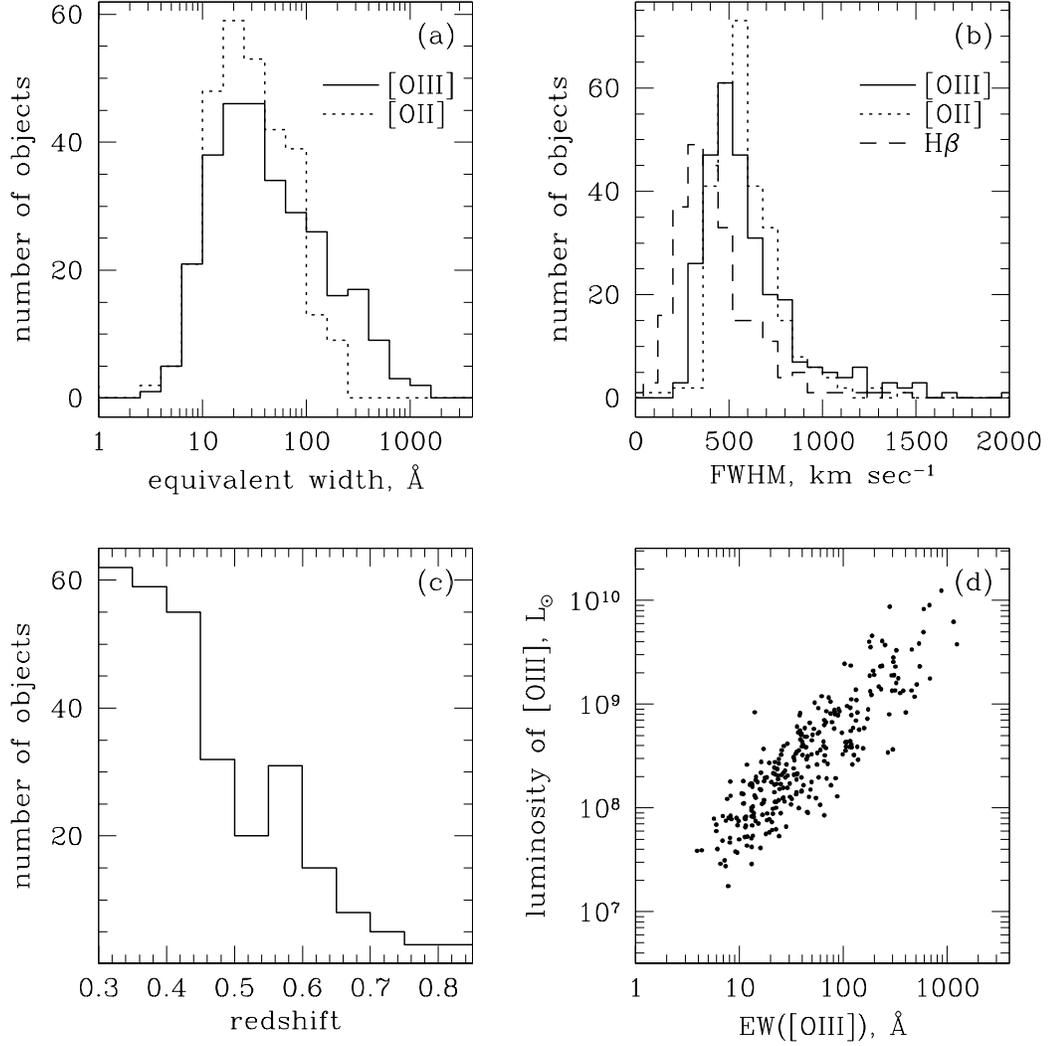}
\figcaption{(a-c): Distribution of the type II AGN sample in the equivalent widths of emission lines ([OIII] 5008 -- solid, [OII] 3728 -- dotted), in the FWHM of emission lines ([OIII] 5008 -- solid, [OII] 3728 -- dotted, H$\beta$ -- dashed) and in redshift. Equivalent widths are calculated relative to the total continuum. (d): Luminosity of the [OIII] 5008 vs its EW. EWs and FWHM are rest-frame values. FWHM are uncorrected for the spectral resolution of $\la 170$ km s$^{-1}$; at the redshifts of the objects in our sample, for a line with intrinsic FWHM$=500$ km s$^{-1}$ this correction is $<3$\%. \label{distribution}}
\end{figure}

Figures \ref{distribution}(a-c) show the distribution of the sample in: (a) equivalent widths of [OII] 3728 and [OIII] 5008 emission lines, (b) FWHM of [OIII], [OII] and H$\beta$, and (c) redshift. EWs are calculated relative to the total continuum (i.e., stellar continuum from the host galaxy included) and corrected to the rest-frame. In Figure \ref{distribution}(b) we show that the permitted lines (H$\beta$) have the same widths or are even narrower than the forbidden lines ([OII] and [OIII]), consistent with the traditional criterion for identifying type II AGN (e.g., \citealt{weed77}). The redshift distribution reflects effects due to spectroscopic targeting and the detectability of type II AGN as strong emission lines move in and out of the $ugriz$ filters. Figure \ref{distribution}(d) shows the luminosity of the [OIII] 5008 line as a function of its EW. The objects from the luminous subsample ($L{\rm [OIII]}>3\times 10^8 L_{\odot}$) typically have very high EWs of the [OIII] 5008 line ($\ga50$\AA).

\begin{figure}
\epsscale{1.0}
\plotone{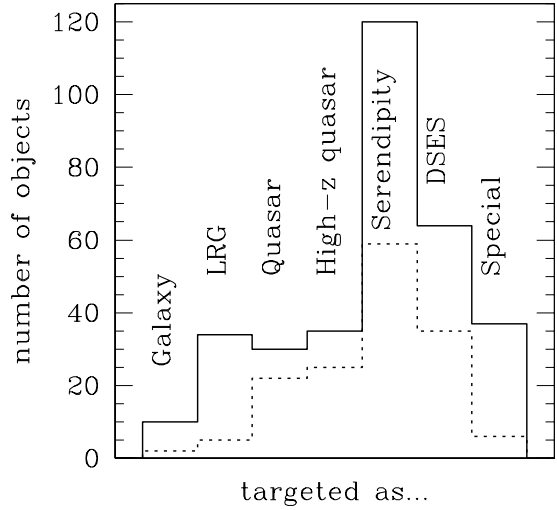}
\figcaption{The distribution in the spectroscopic target selection algorithms for all type II AGN candidates (solid line) and for the high-luminosity subsample $L$[OIII]$>3\times 10^8 L_{\odot}$ (dotted line). An object can appear multiple times on this diagram if it is targeted by several algorithms. \label{target_distribution}}
\end{figure}

Figure \ref{target_distribution} shows the distribution of the objects according to the targeting algorithms. The distribution for the whole sample is shown with a solid line, the distribution for the luminous subsample ($L{\rm [OIII]}>3\times 10^8 L_{\odot}$) is shown with a dotted line. 42\% of the whole sample were targeted only by the Serendipity algorithms, and of these the overwhelming majority (95\%) were targeted as optical counterparts of FIRST sources. 

About 19\% of the objects were observed as part of the DSES survey and about 11\% are from the Special plates. These fractions are quite high considering that DSES and Special plates only constitute 5.5\% and 0.7\%, correspondingly, of the initial sample (33 and 4 plates out of the 600 plates to which the selection algorithm was applied). This suggests that target selection algorithms used for DSES and Special plates favor selection of type II AGN. The high selection efficiency on DSES plates is due to the deeper magnitude limits for the Quasar and Serendipity algorithms on these plates, whereas low-luminosity type II AGN are effectively targeted on Special plates due to the preferential selection of objects bluer than normal galaxies. 

We remark that our sample of type II AGN is not complete. For the samples of low-redshift SDSS AGN by \citet{hao03} and \citet{kauf03} the selection criteria are well-defined because AGN are selected from the main galaxy sample. In our sample, many objects were targeted for spectroscopy as Serendipity sources which are sampled for spectroscopy only if fibers remain available after galaxy and quasar candidates have been allocated fibers. A large fraction of the sample was observed on DSES or Special plates where targeting algorithms varied with time and depended on specific tasks of the specialized surveys. Moreover, spectroscopic selection partly relies on the presence of the weak [NeV] 3347, 3427 lines, and thus is dependent on S/N. 

\subsection{Colors}
\label{colors}

In this Section we compare the colors of type II AGN with those of other objects. For a comparison sample, we selected all objects from 100 spectroscopic plates with assigned redshifts $0.3<Z<0.83$ (about 5700 objects), most of which are LRGs and quasars. 

\begin{figure}
\epsscale{0.9}
\plotone{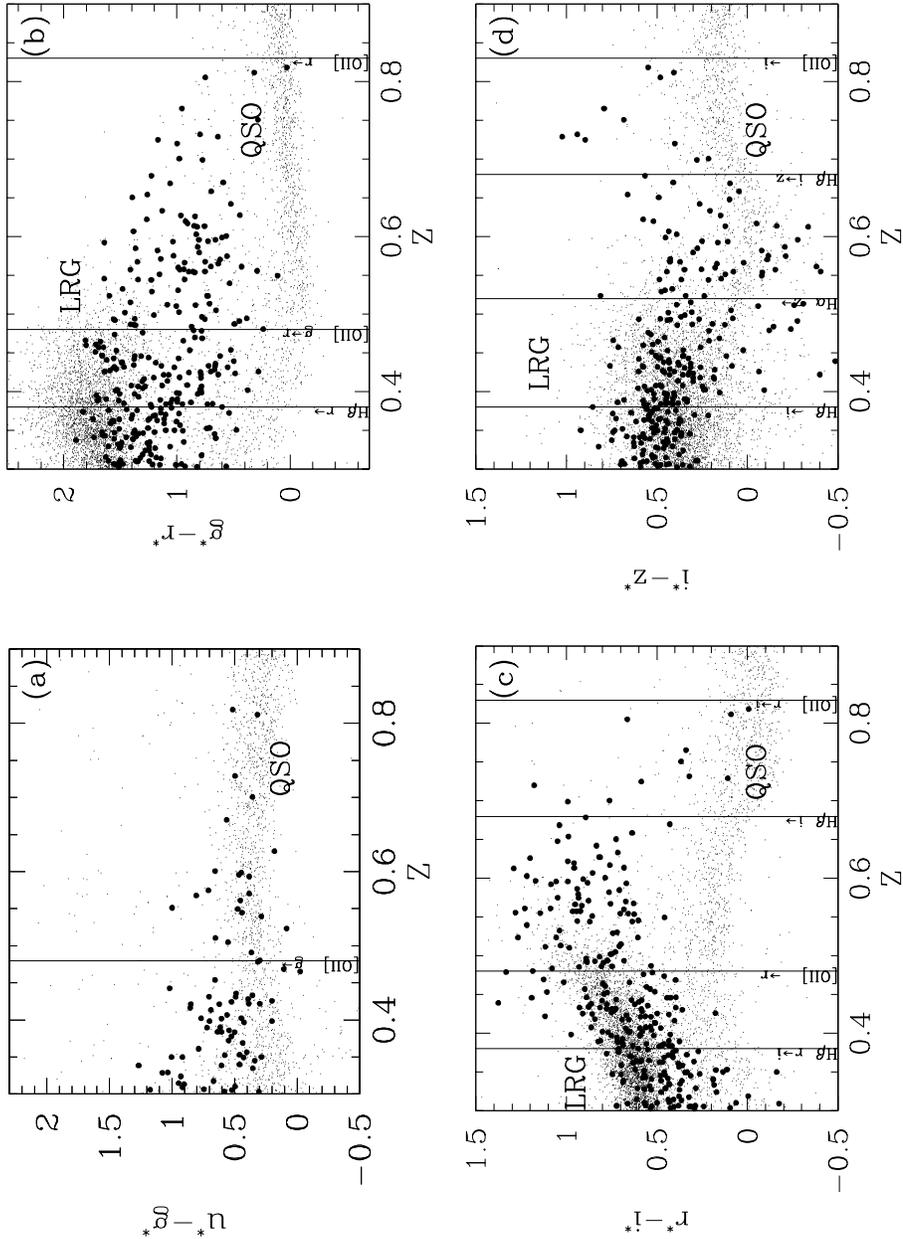}
\figcaption{Redshift-color diagrams: type II AGN (filled circles) and all other SDSS spectroscopic targets at $Z>0.3$, mostly LRGs and quasars (dots). Only the properly detected objects are shown (i.e., with magnitude errors $<0.2$). Since most LRGs are not detected in the $u$ band, the LRG locus is not present in Figure \ref{zc}(a). Vertical lines mark redshifts at which strong emission lines enter and leave filters (the lines and filters involved are indicated). The colors are based on PSF magnitudes for type II AGN and all unresolved spectroscopic objects, and on Petrosian magnitudes for all other objects. The use of PSF magnitudes for type II AGN minimizes contribution from the host galaxies.\label{zc}}
\end{figure}

Figure \ref{zc} shows redshift-color diagrams for type II AGN in comparison with LRGs and quasars. Type II AGN show a large color scatter at a given redshift and generally have colors intermediate between those of LRGs and of quasars. The main quasar locus exhibits complicated redshift-dependent behavior as emission lines move in and out of the filters (see \citealt{rich01} for details); vertical lines indicate redshifts at which color changes are expected to occur. The colors of LRGs behave as expected for an old stellar population placed at different redshifts. 

The range of EWs of bright emission lines accounts for some of the scatter of the colors of type II AGN, as well as for some features of the redshift dependence. For example, the $i^*-z^*$ color in Figure \ref{zc}(d) changes dramatically at $Z=0.68$ when the H$\beta+$[OIII] 4960, 5008 line complex moves from the $i$ band to the $z$ band, similarly to the change of the colors of normal quasars. The maximum equivalent width of the [OIII] 5008 line is about 1400\AA\ for the objects in our sample. Together with the [OIII] 4960 line (its flux is 1/3 of the flux of [OIII] 5008, \citealt{oste89}) and a much weaker H$\beta$, the maximum equivalent width of the line complex is about 1900\AA. The change of magnitude in band $X$ with width $w_X$ (about 1000$\mbox{\AA}$ for SDSS filters) due to the presence of emission lines is
\begin{equation}
\Delta m_X=-2.5 \log ({\rm EW}/w_X)
\end{equation}
and thus the maximum change in color introduced by the H$\beta$+[OIII] complex is about 0.7 mag. The EWs of this line complex range between about 15\AA\ and 1900\AA, and lines with EWs at the lower limit give a negligible contribution to the color. Even if the underlying continuum has the same spectral energy distribution in all objects, the varying EWs of emission lines alone can account for the scatter of the $r^*-i^*$ and $i^*-z^*$ colors of type II AGN in Figures \ref{zc}(c,d). 

The same argument, however, does not apply to Figure \ref{zc}(b), where a large scatter in the $g^*-r^*$ color ($\sim$1.5 mag) is present even though the brightest line complex H$\beta+$[OIII] is redshifted out of both bands at $Z>0.38$. All type II AGN have $g^*-r^*$ color significantly bluer than that of an old stellar population (as can be seen from comparison with the position of the LRG locus). Unlike galaxies, many type II AGN are detected in the $u$ band as we show in Figure \ref{zc}(a), and have colors that are different from those of quasars. In other words, there is evidence for a varying blue component in the spectral energy distribution of type II AGN as compared to an old stellar population.

We have been unable to quantify the contribution of the blue component to the spectra of individual objects. Stellar absorption features (e.g., CaII 3935, 3970 absorption lines) are readily seen in the spectra of many type II AGN, but when we subtract the stellar contribution according to the procedure described in Section \ref{line_width}, the residual continua have very low S/N. We discuss the nature of the blue component in Section \ref{comp_continuum} using the composite spectrum of the entire sample.

\subsection{Emission line properties}
\label{em_lines}

In this section we compare the emission line properties of our sample of type II AGN with those of broad-line AGN. For a comparison sample, we took all objects from 100 plates with assigned redshifts $Z<0.83$ spectroscopically classified as broad-line AGN -- overall, 1954 objects. We verified the automated spectral classifications by visually inspecting all spectra from the comparison sample. Due to the redshift constraint, the [OIII] 5008 line region is covered in all spectra.

\begin{figure}
\epsscale{0.9}
\plotone{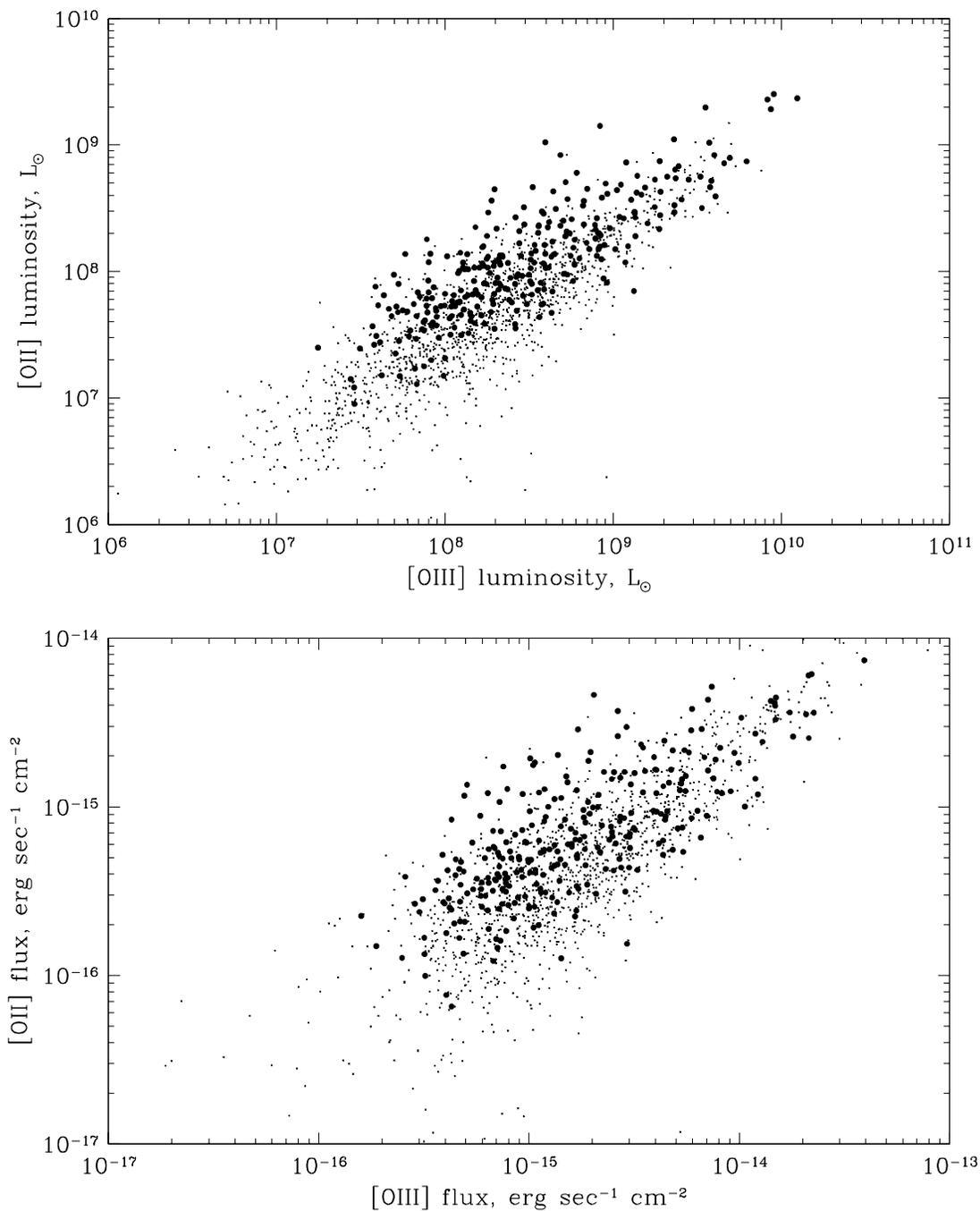}
\figcaption{Luminosities (top) and fluxes (bottom) of the [OII] 3728 and [OIII] 5008 lines: SDSS broad-line AGN (small symbols) and type II AGN (circles). The luminosities are well-correlated for all AGN and for type II AGN in particular. Luminosities are given in solar luminosities $L_{\odot}$ and fluxes are in erg s$^{-1}$ cm$^{-2}$.\label{oii_oiii}}  
\end{figure}

Figure \ref{oii_oiii} shows the fluxes and luminosities of the [OII] 3728 and [OIII] 5008 lines for type II AGN (big filled circles) and for type I AGN from the comparison sample (small symbols). We obtain fluxes and luminosities after fitting the lines with Gaussian profiles. The linear fit to the data for broad-line AGN yields
\begin{equation}
\log L{\rm [OII]}=(1.1\pm 0.3)\log L{\rm [OIII]}+(-1.4\pm 2.5),
\end{equation}
with correlation coefficient $r^2=0.61$. For 90\% of broad-line AGN, the range of line luminosity ratios is $L$[OIII]/$L$[OII]$=1.5$--10, with 4 being the median value, consistent with the ratio of line fluxes of 5.97 for the composite spectrum of quasars  \citep{vand01} and with other studies (e.g., \citealt{kura00} and references therein). 

\begin{figure}
\epsscale{1.0}
\plotone{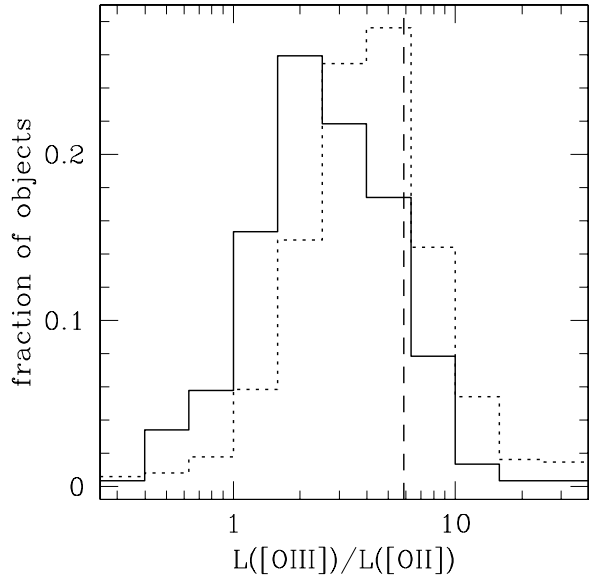}
\figcaption{Distribution of the [OIII] 5008/[OII] 3728 ratio for broad-line AGN (dotted line) and type II AGN (solid line). For a given [OII] 3728 luminosity, [OIII] 5008 lines appear slightly fainter in type II AGN. The vertical dashed line marks the [OIII] 5008/[OII] 3728 ratio for broad-line quasars from the composite spectrum by \citet{vand01}.\label{oii_oiii_distribution}}   
\end{figure}

In Figure \ref{oii_oiii}, type II AGN tend on average to have slightly less luminous [OIII] 5008 for a given [OII] 3728 luminosity than broad-line AGN. To quantify this trend we plot the distributions of [OIII]/[OII] ratios for both types of AGN in Figure \ref{oii_oiii_distribution}. The effect is opposite in sign to reddening: the shorter-wavelength line would have been extincted more in obscured AGN if the two lines originated in the same location. Our result is in agreement with many previous studies showing that [OIII] 5008/[OII] 3728 ratios are higher in broad-line AGN than in narrow-line AGN \citep{saun89, jack90, tadh93}. 

This trend might imply that the ionization parameter of the narrow-line region (the ratio of ionizing flux to density) that governs the balance of ionized species is higher in type I AGN than in type II AGN. This difference could be due to geometric effects related to anisotropic obscuration. Alternatively, one might suggest that the [OIII]-emitting region is subject to more reddening than the [OII]-emitting region. In this case a necessary requirement is that the ionization parameter should rise inward. The contribution to line emission from the star-formation in the host galaxy can also affect line ratios. More detailed studies of several emission lines would be required to distinguish between these possibilities. S/N of individual objects in our sample does not allow us to study other emission lines; we compare line ratios of type II AGN to those of type I AGN in Section \ref{comp_em_lines} using the composite spectrum of type II AGN.

\subsection{Objects with unusual line profiles}
\label{unusual}

The shapes of the emission lines in quasars and Seyfert galaxies have long been known to deviate from Gaussianity. For example, \citet{boro92} describe how the shape of the H$\beta$ line is correlated with other emission line properties and with the broad-band characteristics of quasars. The shape of the [OIII] 5008 emission line was studied for large samples of AGN by \citet{heck81} and \citet{whit85}; this line was often found to show significant asymmetries.

Visual inspection of our sample of type II AGN revealed 31 objects with [OIII] 5008 line profiles that deviate substantially from Gaussian (i.e., they show significant asymmetries and/or substructure). We will address these objects in greater detail in our future work, but we present here a qualitative description of this subsample. Objects discussed in this Section are marked with the flag `asym' in the column ``comments'' of Table 1.

\begin{figure}
\epsscale{0.9}
\plotone{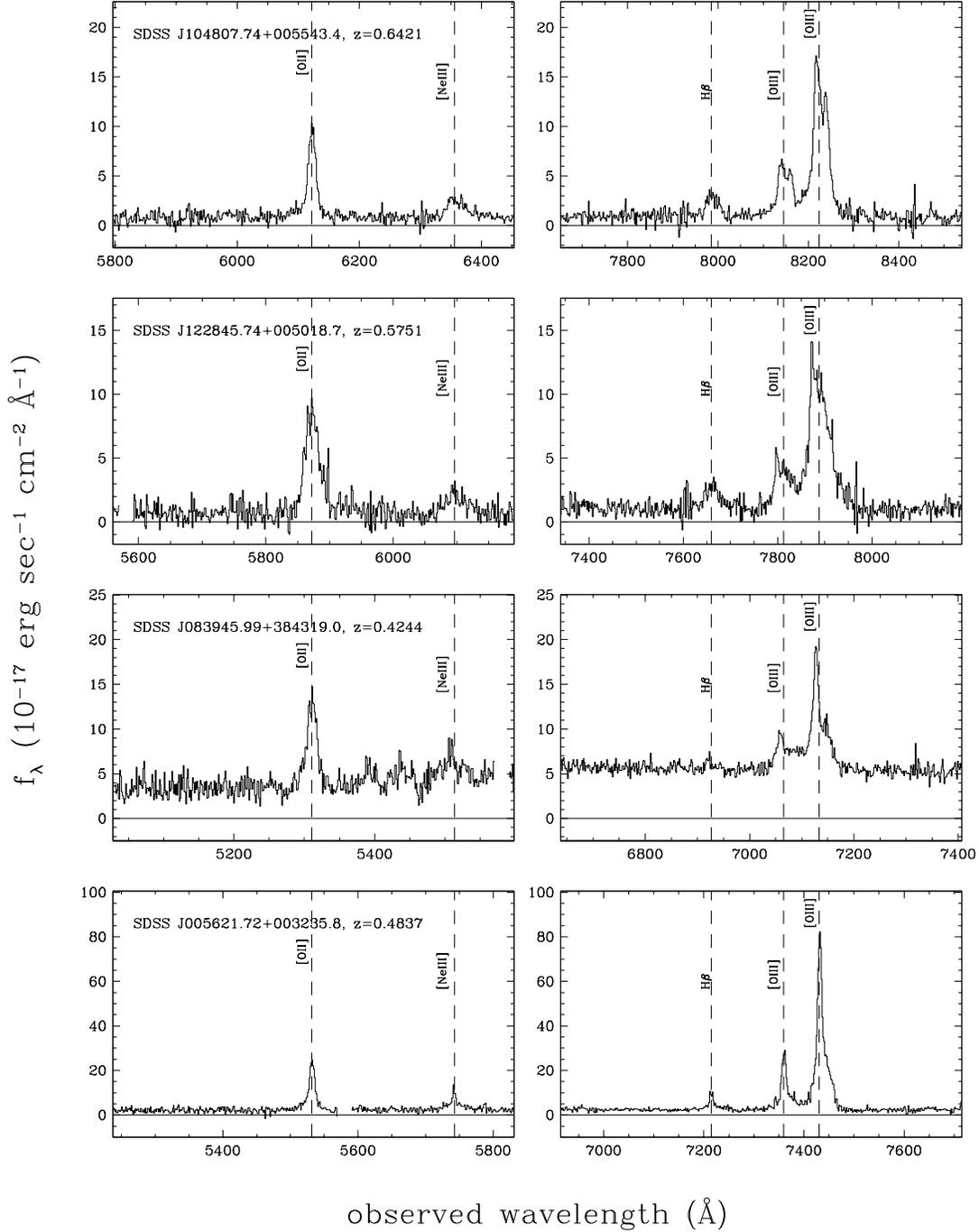}
\figcaption{Example spectra of objects with red asymmetries in [OIII] 5008 smoothed by 5 pixels. [OII] 3728 and [NeIII] 3870 are shown in the left panels and H$\beta$+[OIII] 4960, 5008 in the right panels. Vertical dashed lines indicate where the emission lines should be if we adopt the centroid redshift of the [OII] 3728 lines, as determined by fitting [OII] 3728 with a single Gaussian profile. \label{split}}
\end{figure}

Figure \ref{split} gives example spectra of objects with red asymmetries in the [OIII] 5008 emission line, i.e., in which more flux is emitted redward of the peak of the emission line than blueward. We find five such objects (out of 31 objects with significant asymmetries), and four of these show two distinct components in the line profile, similarly to SDSS J104807.74$+$005543.4 (Figure \ref{split} top). A typical splitting between the two peaks is $\sim$ 700 km s$^{-1}$, as in the case of SDSS J104807.74$+$005543.4. The last object of the red-asymmetric group, SDSS J005621.72$+$003235.8 (Figure \ref{split} bottom) does not show a second peak, but rather shows a smooth profile.

Another three objects (SDSS J135128.14$-$001016.9, SDSS J225612.18$-$010508.1 and SDSS J225227.39$-$005528.5) show a double-horned [OIII] 5008 profile, with the two components being approximately equal in strength.

\begin{figure}
\epsscale{0.9}
\plotone{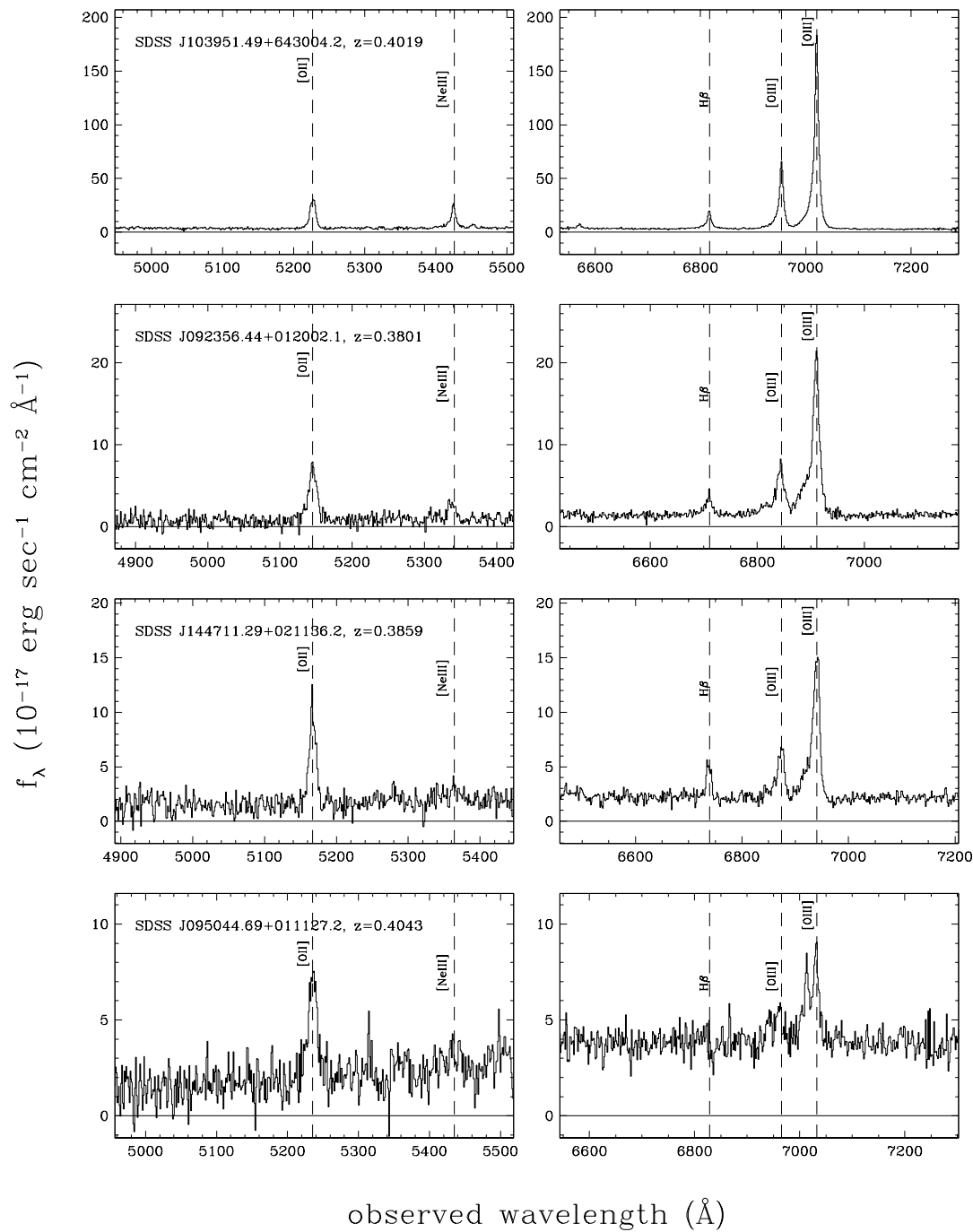}
\figcaption{Same as Figure \ref{split}, but for objects with blue asymmetries in [OIII] 5008.\label{asym}}
\end{figure}

Figure \ref{asym} gives example spectra of objects with blue asymmetries. We find 23 such objects, but only one of them shows split lines (SDSS J095044.69$+$011127.2; Figure \ref{asym} bottom). Most of the objects with blue asymmetries show smooth [OIII] 5008 profile, similar to SDSS J103951.49$+$643004.2 (Figure \ref{asym} top).

In all cases, the [OII] 3728 was well detected, and we did not find any significant asymmetries in this line. Other emission lines are detected with much lower S/N. We do not see any evidence for irregularities in the shape of H$\beta$, but in some cases [NeIII] 3870 has an asymmetric profile similar to that of the [OIII] 5008 line (e.g., SDSS J104807.74$+$005543.4; Figure \ref{split} top).  

\section{Composite spectrum}
\label{composite}

We produced a composite spectrum of the type II AGN in order to reveal weak emission lines, to study the underlying continuum, and to discuss possible types of contaminants to the sample.

\subsection{Generating the composites}

We start by shifting each spectrum to the rest frame determined from the [OIII] 5008 emission line redshift. Constructing the composite using the redshift of the [OII] 3728 line made no significant difference. The spectra were then rebinned onto a common wavelength scale at 1$\mbox{\AA}$ per bin.

We produced two composite spectra using different combining (weighting) techniques: 1) variance-weighted combining, which preserves the relative fluxes of the emission lines; 2) geometric mean combining, which preserves the global continuum shape \citep{vand01}. The resulting spectra are presented in Figure \ref{two_composites} (a) and (b), respectively. Since fewer objects contribute to the ends of the covered spectral range, the blue and red parts of the composite spectra have lower S/N than the central part. 

\subsection{Emission line ratios}
\label{comp_em_lines}

\begin{figure}
\epsscale{0.9}
\plotone{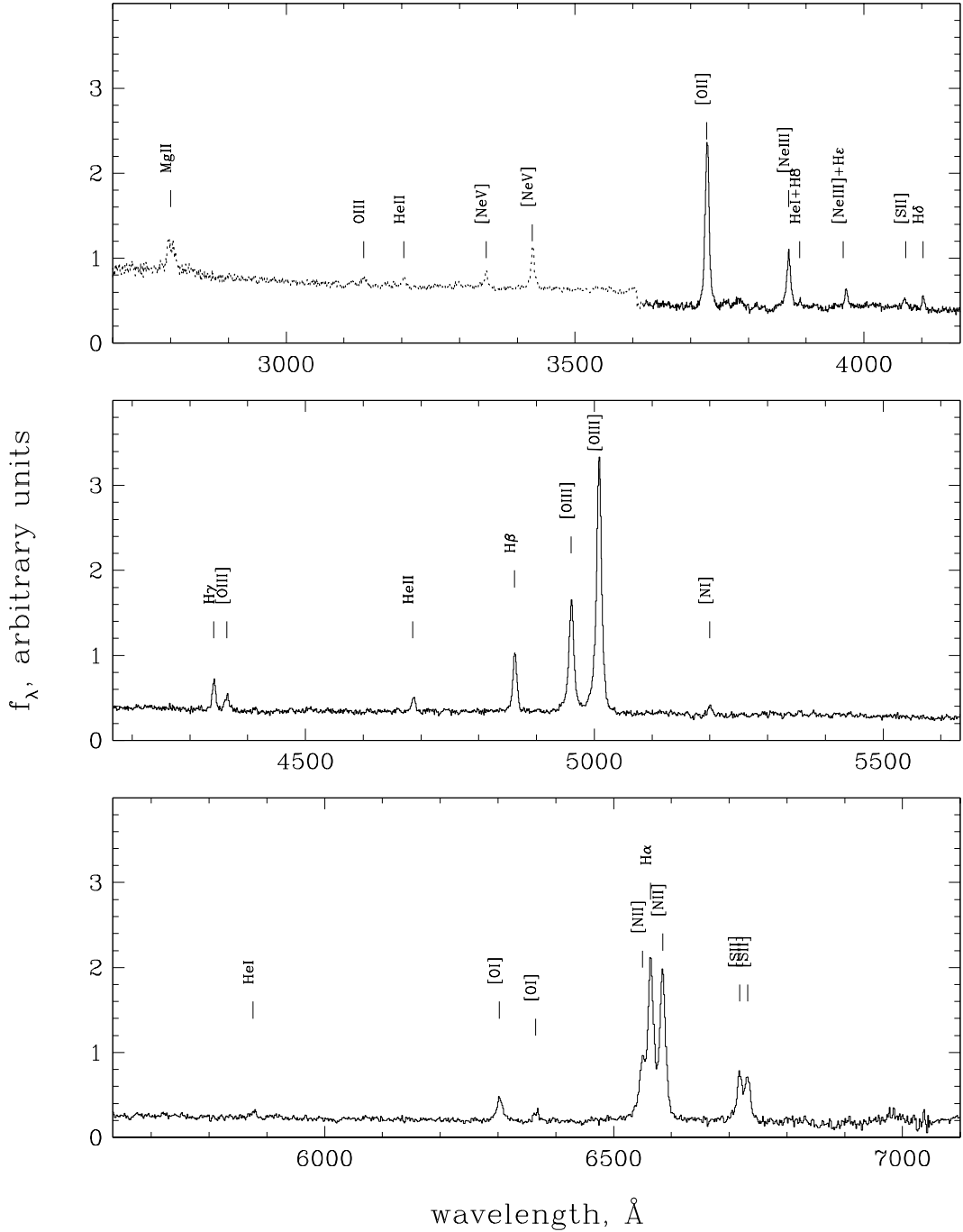}
\figcaption{Expanded view of the variance-weighted composite spectrum of type II AGN with all detected emission lines marked. The host galaxy contribution is subtracted away to reveal weak emission lines. The galaxy templates do not extend blueward of $\lambda=3610\mbox{\AA}$; we plot the unsubtracted part of the spectrum with a dotted line. The spectrum is not smoothed.\label{blowup}}
\end{figure}

An expanded version of the variance-weighted composite spectrum in the range 2700-7100 \AA\ is shown in Figure \ref{blowup}. The contribution from the host galaxy was subtracted according to the procedure described in Section \ref{line_width}. The part of the spectrum where host galaxy templates are unavailable is plotted with a dotted line. We compared the composite spectrum of type II AGN to the composite spectrum of SDSS quasars \citep{vand01} and identified all emission features as labeled in Figure \ref{blowup} based on their wavelength position (see \citealt{vand01} for a complete list of references on line identifications). In this manner we identified 26 emission features in this wavelength range. 

We then fit the emission lines with Gaussian profiles, and from the best-fit parameters of the Gaussian fitting we calculate relative fluxes of emission lines. Emission line parameters are listed in Table 2. Single Gaussian fitting was used for most lines. In order to perform the fitting of H$\alpha+$[NII], we assume that the widths of the [NII] 6550, 6585 lines are the same and that their flux ratio is 1/3 \citep{oste89}. The MgII 2796, 2804 doublet is resolved in the composite spectrum, and we fit it with two Gaussian components of the same width. The relative line fluxes are found to be in good agreement with those of a composite spectrum of Seyfert II galaxies from \citet{oste93}. In table 2, we give relative fluxes of forbidden lines from the composite spectrum of broad-line SDSS quasars \citep{vand01}.

Comparing the relative line fluxes of type II AGN and relative fluxes of forbidden lines of broad-line quasars in Table 2, we find that high-ionization lines ([OIII] 4960, 5008; [NeIII] 3870; [NeV] 3347, 3427) tend to be weaker in type II AGN than in quasars, consistent with \citet{naga01} who studied the dependence of ionization ratios on the Seyfert type. This effect is particularly strong for the highest-ionization lines [NeV] 3347, 3427. For individual objects, the same effect is seen for the high S/N emission line [OIII] 5008 (Section \ref{em_lines}). 

The H$\alpha$/H$\beta$ ratio of 4.0 in the type II AGN composite suggests that there is a reddening of about $E(B-V)=0.27$ mag toward the region where the narrow Balmer lines originate, assuming that the intrinsic ratio is (H$\alpha$/H$\beta$)$_0$=3.1 (e.g., \citealt{oste93}). 

\subsection{Possible contaminants}
\label{comp_contaminants}

\begin{figure}
\epsscale{1.0}
\plotone{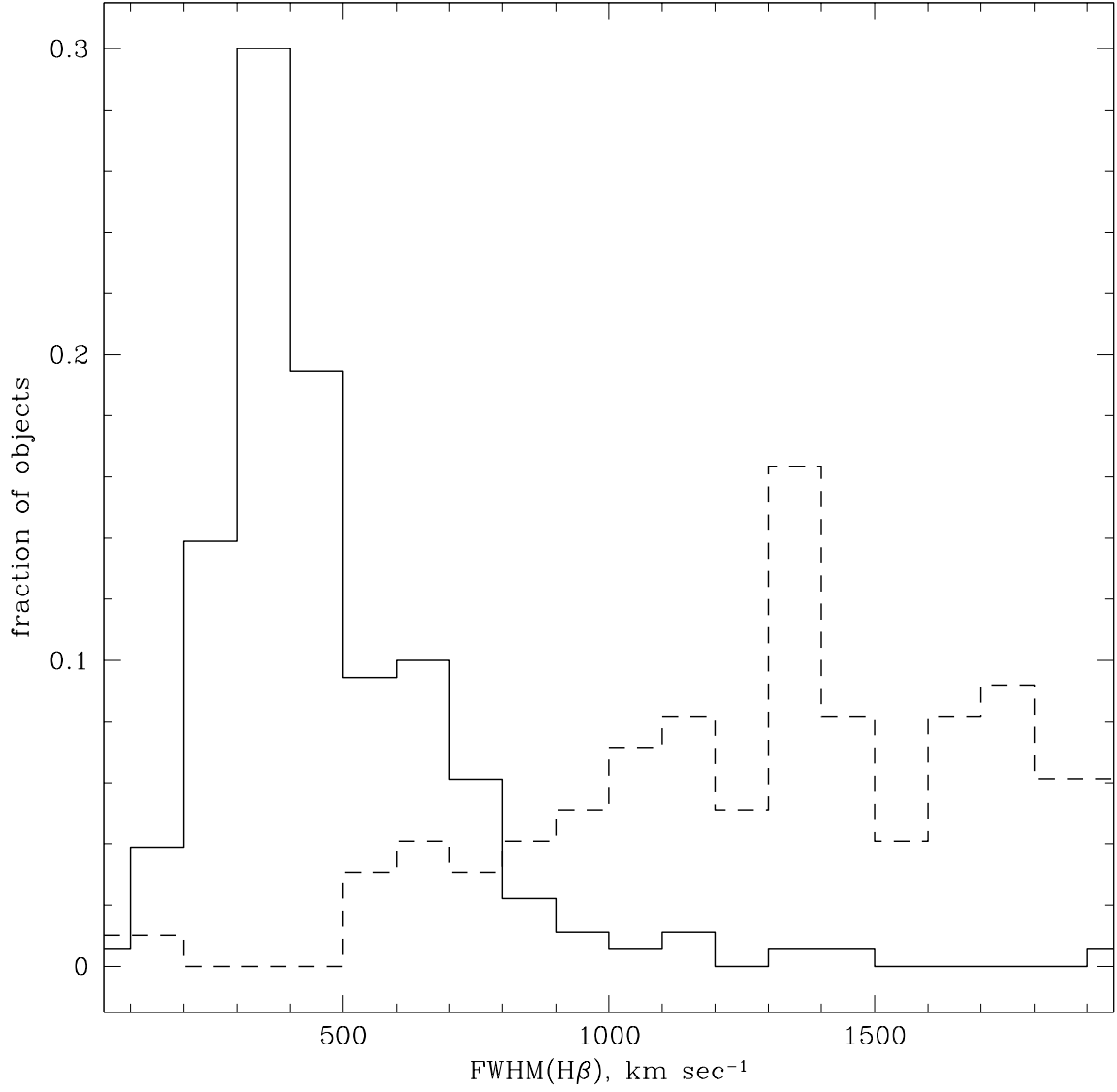}
\figcaption{The distribution of FWHM(H$\beta$) for type II AGN is shown with the solid line. The distribution of FWHM(H$\beta$) for NLSy1 from \citet{will02} is shown with the dashed line.\label{hbeta_width}}
\end{figure}

The variance-weighted composite spectrum presented in Figure \ref{two_composites}(a) shows no sign of the FeII emission blend around $4750\mbox{\AA}$ prominent in high S/N spectra of NLSy1 \citep{halp98}. In addition, it is clear from Figure \ref{hbeta_width} that the distributions of H$\beta$ FWHM are very different for type II AGN and for NLSy1 from \citet{will02}. Furthermore, as discussed in Section \ref{selection}, our high-ionization criteria strongly disfavor selection of NLSy1, and the FeII emission blend is not seen in the spectra of any individual objects in our sample. This suggests that our sample is not contaminated by appreciable numbers of NLSy1.

We also do not see any evidence for broad components in permitted lines (MgII 2800, H$\alpha$, H$\beta$). We conclude that our selection procedure efficiently rejected objects with weak underlying broad components in these lines.

\subsection{Continuum}
\label{comp_continuum}

In Section \ref{colors}, based on the photometric data, we found that type II AGN are bluer in $u^*-g^*$ and $g^*-r^*$ colors than early-type galaxies, and emission lines are not sufficient to explain this color difference. The continuum flux density of type II AGN in our sample is for the most part detected with low S/N and the shape of the continuum cannot be studied based on spectra of individual objects. The contributions to the continuum can now be discussed using the high S/N composite spectrum.

\begin{figure}
\epsscale{0.9}
\plotone{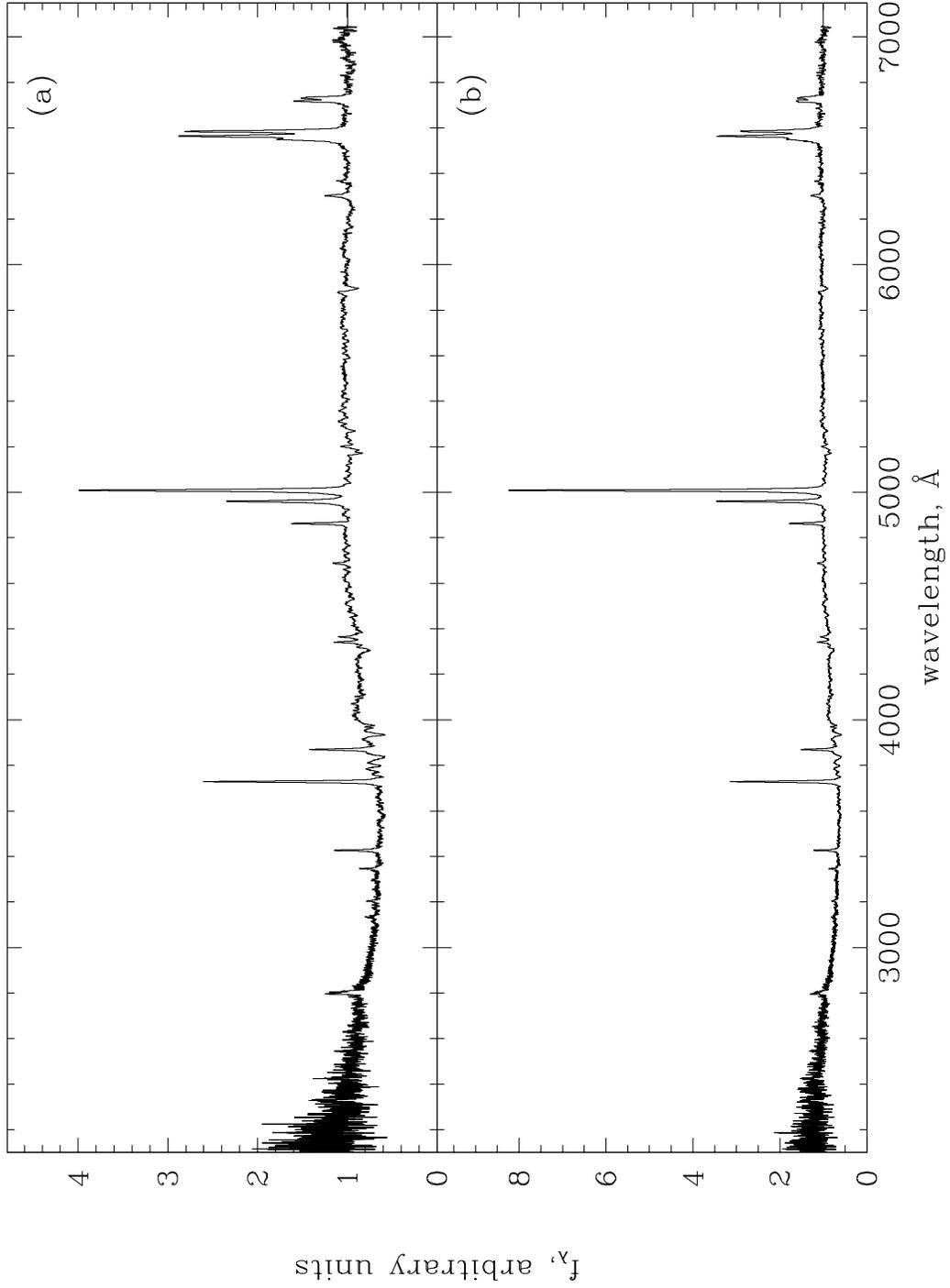}
\figcaption{The composite spectra of type II AGN: (a) based on variance-weighted combining; (b) based on geometric mean combining (see text for details). The spectra are not smoothed and are scaled to have the median $f_{\lambda}=1$.\label{two_composites}}
\end{figure}

The composite spectra (Figure \ref{two_composites}) show evidence for an old stellar population through CaII 3935, 3970 absorption lines. Apart from this contribution, there is evidence for the presence of young stars through the high-order Balmer absorption lines (four absorption features, H9 -- H12, between the [OII] 3728 and the [NeIII] 3870 emission lines are clearly detected) and the strength of the Balmer edge at 3650\AA. These features are very characteristic of A stars, and are also the strongest spectral features of O and B stars in the optical. 

A-type stars, however, cannot explain the strong blue and UV continuum apparent in Figure \ref{two_composites}. The problem of the excess blue continuum of Seyfert II galaxies has been known for a long time (e.g., \citealt{anto93}). A number of options have been discussed (see \citealt{cid95} and \citealt{heck97}) and some have been convincingly ruled out. For example, thermal emission from the scattering plasma \citep{mill94} produces a very high ratio of HeII 4687 to H$\beta$. Our HeII 4687/H$\beta$ of $\sim0.2$ is consistent with predictions of standard photoionization models \citep{ferl86} and, following the argument by \citet{heck97}, we conclude that the contribution to the continuum from this mechanism is negligible.

Scattered light from the nucleus is another possibility but the broad components of permitted emission lines would have been scattered, too. Since broad components are not observed, an upper limit can be placed on the contribution from the scattered light from the nucleus. Assuming that the scattered component has the same spectrum as the composite spectrum of broad-line SDSS quasars \citep{vand01}, we estimate that no more than 8\% of the continuum of type II AGN at $5000\mbox{\AA}$ and no more than 20\% of the continuum at $3000\mbox{\AA}$ can be due to the scattered light. 

The resolution of the problem of excess blue continuum appears to be the strong correlation between nuclear activity and starbursts in the host galaxy \citep{gonz93, heck95, cid01, kauf03}. Very young O and B stars can supply the necessary amount of UV continuum, their only observable spectroscopic features in the optical being Balmer absorption lines and the Balmer edge.

\begin{figure}
\epsscale{0.9}
\plotone{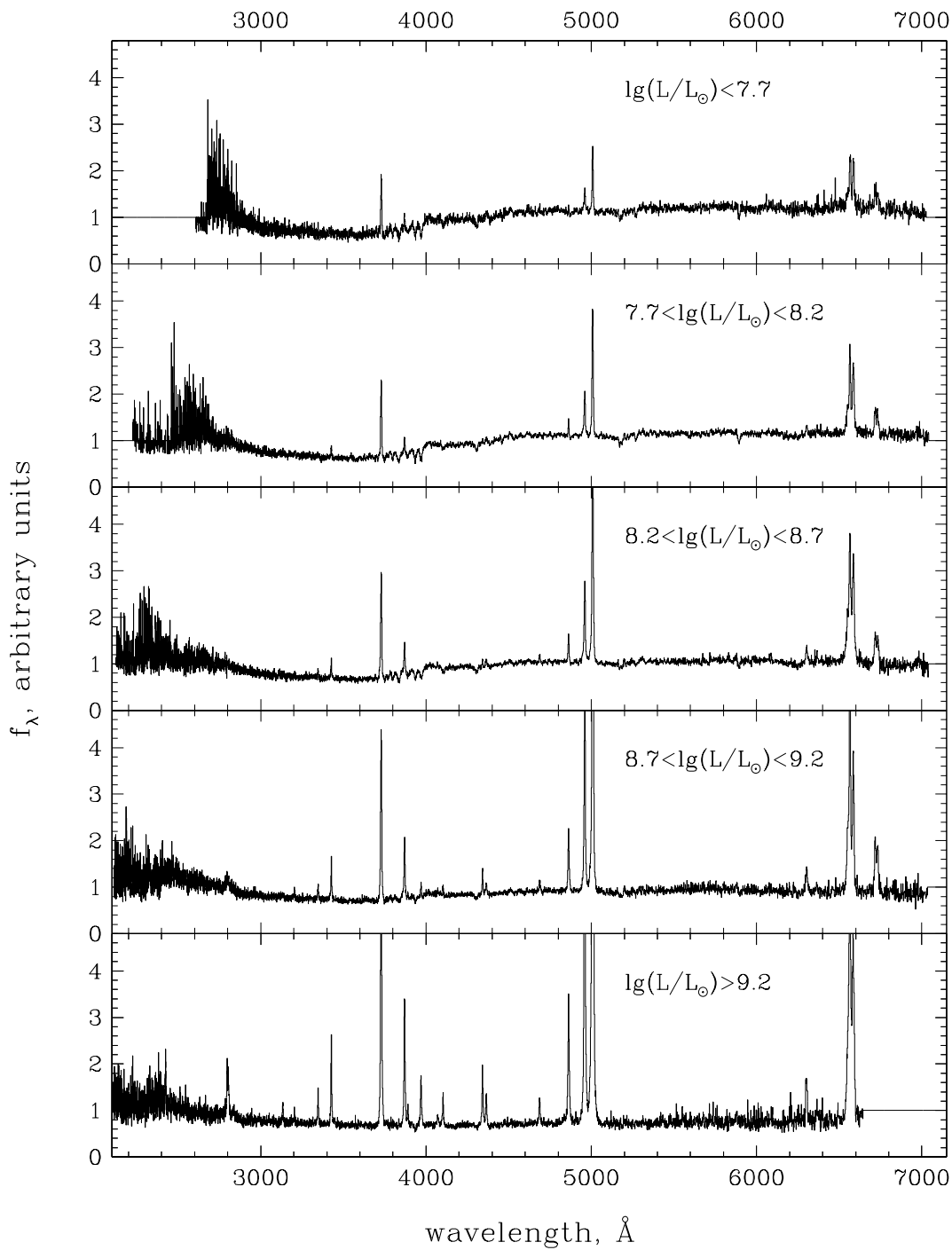}
\figcaption{Geometric mean composite spectra for samples in bins of [OIII] 5008 line luminosity. $L$[OIII] increases downward. At least 23 objects contributed to each bin. The spectra are not smoothed and are scaled to have the median $f_{\lambda}=1$. \label{binned}}
\end{figure}

To study this possibility further, in Figure \ref{binned} we plot geometric mean composite spectra in logarithmic bins of the [OIII] 5008 line luminosities. Unexpectedly, with increasing line luminosity, Balmer absorption lines and Balmer edge disappear. In fact, no stellar features from young stars are apparent in the two highest luminosity bins (the CaII 3970 absorption line is due to the old stellar population). The absence of these features had been historically put in an argument against the stellar origin of the blue continuum. We remark that the highest-luminosity composite spectrum looks similar to the optical spectrum of Mrk477, a nearby luminous type II AGN, which also does not have absorption features from young stars. In Mrk477 a starburst is directly detected in the far-UV, and \citet{heck97} showed that Balmer absorption lines and Balmer edge due to O and B stars are washed out by line emission in this object. We suggest that the same mechanism is operating in the most luminous type II AGN in our sample.

As was mentioned before, the SDSS fibers subtend more than 13.4 kpc on the sky at the redshifts of the objects in the sample, and the spectra include most of the light from the host galaxy, so from the spectra alone we cannot infer whether the starburst is circumnuclear or occurs in the outer parts of the galaxy. Traditionally it was thought that the starburst occurs on the sub-kpc scales near the nucleus; this is supported by investigations of nearby Seyfert II galaxies showing circumnuclear starbursts with sizes of several hundred parsecs \citep{gonz01}. However, \citet{kauf03} argued recently that the starbursts in AGN are spread out over several kpc.

\section{Discussion: Intrinsic luminosities}
\label{discussion}

As discussed in Section \ref{introduction}, multi-wavelength observations are essential to develop a physical model for type II AGN. In particular, if these objects are obscured, the optical continuum magnitude is by definition a poor indication of the intrinsic luminosity. We therefore use the [OIII] 5008 emission line as a proxy of the AGN activity, as it is emitted from the extended (and therefore presumably less obscured) narrow line region. 

There is no general agreement in the literature on which narrow emission line, [OIII] 5008 or [OII] 3728, can best serve as a nuclear luminosity indicator. In this work and in many previous studies (Section \ref{em_lines} and references therein) it was concluded that the [OIII] 5008/[OII] 3728 ratios are higher in broad-line AGN than in narrow-line AGN, but \citet{kura00} do not confirm this trend. \citet{simp98} argues that such studies suffer from incompleteness and that the [OIII] 5008 line is a good indicator of the underlying nuclear continuum. \citet{croo02} find that [OII] 3728 can be significantly contaminated by star formation in the host galaxy, while \citet{kauf03} suggest that the same effect is negligible for the [OIII] 5008 line. Following these authors, we use [OIII] 5008 as the luminosity indicator.

In addition to all these possible complications, all emission lines can be subject to extinction by interstellar dust in the host galaxy and in our Galaxy. This effect is usually corrected for by using the Balmer decrement, but we do not have spectral coverage of the H$\alpha$ line for most of the sample. Interstellar extinction should affect type I and type II AGN in a similar way, so we ignore this effect for the purposes of the comparative analysis.

\subsection{[OIII] 5008 vs broad-band luminosity in broad-line AGN} 

The luminosity of the [OIII] 5008 line correlates with numerous multi-wavelength broad-band characteristics of type I (unobscured) AGN (e.g., \citealt{mulc94}). In the optical, a strong correlation has been found between the strength of [OIII] 5008 and the primary eigenvector derived from the PCA of luminous quasars \citep{boro92}. Below we discuss a correlation between the [OIII] 5008 luminosity and the continuum luminosity as characterized by rest-frame $B$-band absolute magnitude. 

We study the sample of 1954 SDSS broad-line AGN at $Z<0.83$ used in Section \ref{em_lines}. To obtain a measure of broad-band optical luminosity, we calculate the total luminosity emitted between rest-frame wavelengths $\lambda_{\rm start}=3980\mbox{\AA}$ and $\lambda_{\rm end}=4920\mbox{\AA}$. This spectral region conveniently avoids bright emission lines and has the advantage of being very close in wavelength coverage to the conventionally used $B$-band. The observed flux of the rest-frame $B$-band is then
\begin{equation}
F_B=\int_{\lambda_{\rm start}(1+z)}^{\lambda_{\rm end}(1+z)}\,f(\lambda){\rm d}\lambda.
\end{equation}
We define the $B$-band luminosity and the absolute magnitude as:
\begin{equation}
L_B=4\pi D_L^2 F_B; \quad
M_B=-2.5 \log \left(F_B\frac{(D_L)^2}{(10{\rm pc})^2}\right) -13.04, \label{MB}
\end{equation}
where $D_L$ is the luminosity distance to the object. The numerical term in the second definition is taken to match the conventionally used $M_B$ magnitude (e.g., \citealt{binn98}), if $F_B$ is in units of erg s$^{-1}$ cm$^{-2}$. For emission lines, we use their fluxes and luminosities as calculated from the Gaussian fitting.

\begin{figure}
\epsscale{0.9}
\plotone{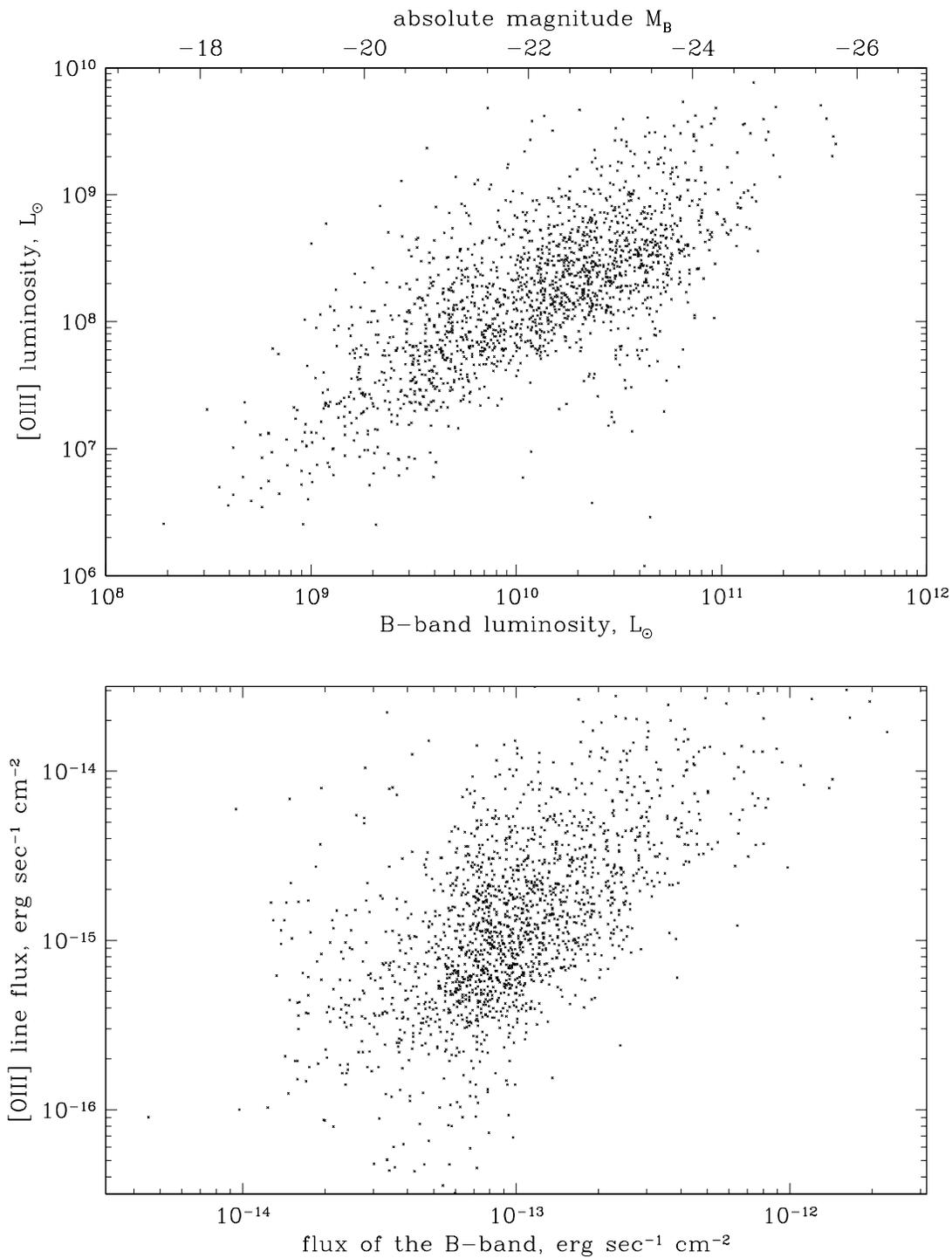}
\figcaption{Top: [OIII] 5008 luminosities versus rest-frame $B$-band luminosities and absolute magnitudes as defined in equation (\ref{MB}) for broad-line AGN. Luminosities are given in solar luminosities $L_{\odot}$. Bottom: observed [OIII] 5008 flux versus the observed flux from the rest-frame $B$-band. Fluxes are in units of erg s$^{-1}$ cm$^{-2}$. \label{line_continuum}}
\end{figure}

Figure \ref{line_continuum} shows the [OIII] 5008 line luminosities vs the rest-frame $B$-band luminosities for SDSS broad-line AGN. The best linear fit to the data points is 
\begin{equation}
\log L{\rm [OIII]}=(1.15\pm0.40)\log L_B+(-3.4\pm 4.0),\label{magn}
\end{equation}
with correlation coefficient $r^2=0.48$. Although the scatter is large, for 75\% of broad-line AGN the range of luminosity ratios is $L_B$/$L$[OIII]$=$30--200, with 100 being the median value.

Within the type I AGN class, the separation between Seyfert 1 galaxies and quasars is somewhat arbitrary, with quasars representing the high-luminosity end of the luminosity function of unobscured AGN. We follow the criterion of \citet{schm83}: an object is defined to be a quasar rather than a Seyfert 1 galaxy if the absolute rest-frame $B$-band magnitude is $M_B<-23$. (We use the $\Lambda$-cosmology throughout this paper. The magnitude cut of $M_B=-23$ in the $\Lambda$-cosmology corresponds to $M_B=-23.46$ at $Z=0.3$ and to $M_B=-23.20$ at $Z=0.8$ in a cosmology with $h=0.5$, $\Omega_m=1$, $\Omega_{\Lambda}=0$ which has been widely used in the AGN literature. The difference between absolute magnitudes in these two cosmologies is described in detail in \citealt{schn03}.) According to equation (\ref{MB}), the magnitude cut of $M_B<-23$ corresponds to the luminosity cut of $L_B>1.2\times 10^{44}$ erg s$^{-1}=2.9\times 10^{10} L_{\odot}$.

\subsection{Estimating the unobscured broad-band luminosities of type II AGN} 
\label{3108}

\begin{figure}
\epsscale{1.0}
\plotone{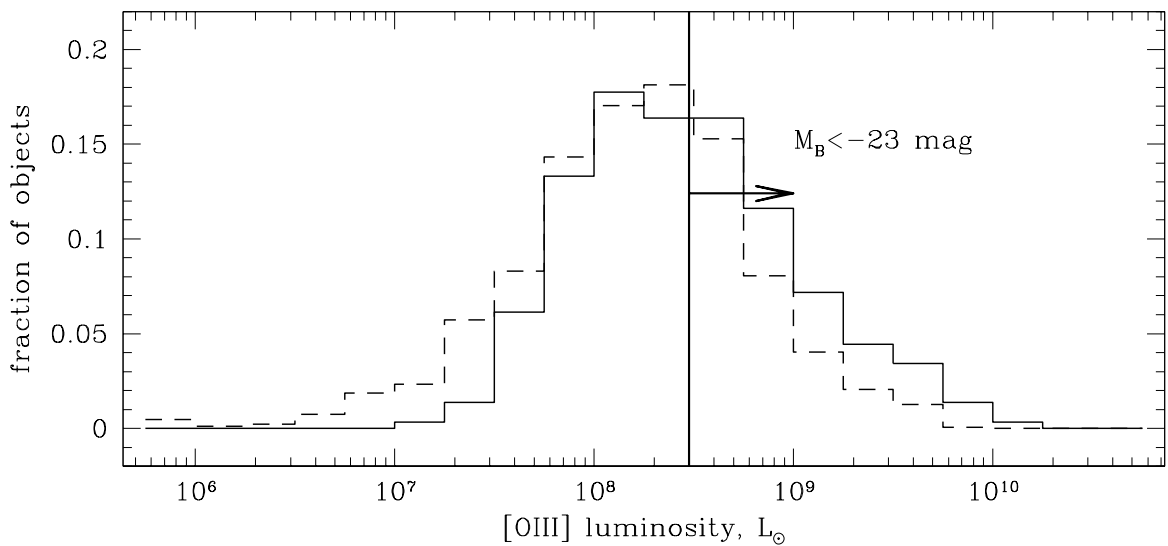}
\figcaption{Number distribution of the [OIII] 5008 line luminosity of type II AGN (solid histogram) in comparison with a sample of broad-line AGN (dashed histogram). On the right of the vertical line lie the broad-line AGN with broad-band luminosities brighter than $M_B=-23$. About 50\% of type II AGN in our sample have [OIII] 5008 luminosities comparable to those of luminous quasars.\label{oiii_lum_dist}}
\end{figure}

We now compare the [OIII] 5008 line luminosities of type II AGN with those of broad-line AGN in Figure \ref{oiii_lum_dist}. Although both samples suffer from incompleteness, we find that about 50\% of type II AGN have [OIII] 5008 luminosities comparable to those of luminous quasars defined as having $M_B<-23$ (the corresponding [OIII] 5008 luminosity is $> 3\times 10^8 L_{\odot}$). Under the assumptions that the narrow-line region is illuminated by the (unseen) photoionizing source the same way in type II AGN as in Seyfert I galaxies and in quasars and that the interstellar extinction affects narrow line regions of both types of AGN in similar ways, we conclude that about 50\% of the objects in our sample have unobscured absolute magnitudes (i.e. as seen if all obscuring material were removed from the line of sight) $M_B<-23$. 

Dereddening has been discussed as an alternative way to estimate obscuration even for heavily reddened objects (e.g., \citealt{hall02}). In type II AGN visual extinction reaches hundreds of magnitudes, and direct emission from the nucleus is unobservable, so dereddening cannot be applied. 

Finally, if the [OIII] 5008 line is partly obscured in type II AGN by circumnuclear material as discussed above, then by using the [OIII]/$M_B$ correlation we underestimate their unobscured luminosities.

\section{Conclusions}
\label{conclusions}

Numerous type II quasar candidates have been found using the SDSS, demonstrating that they are not exceedingly rare. Selecting such a large number of these optically faint objects is possible because of the extensive size of the spectroscopic survey, part of which is dedicated to the search for unusual objects. 

In Section \ref{selection} we described the details of the selection procedure. We looked for high equivalent width, narrow emission line objects with high-ionization line ratios in the redshift range $0.3<Z<0.83$. In particular, we generalized previously developed high-ionization criteria to distinguish type II AGN from other emission-line objects at high redshifts ($Z>0.4$). We found that narrow-line Seyfert 1 galaxies and low-metallicity star-forming galaxies are successfully rejected from the sample by our selection criteria.

We described the sample and its optical properties in Section \ref{properties}. The sample consists of 291 objects, representing $\sim 0.08\%$ of all SDSS spectra or $\sim 2\%$ of all AGN spectra in the same redshift range. These objects are difficult to recover in other optical surveys because they are optically faint, the median magnitudes of the sample being $\left<g^*\right>=21.5$, $\left<r^*\right>=20.4$ and $\left<i^*\right>=19.8$. Unlike radio galaxies, most of the objects in our sample are radio-quiet, as we will demonstrate in the follow-up work (Zakamska et al., 2003, in preparation).

We presented a subsample of 31 objects with unusual line profiles (split or asymmetric) in Section \ref{unusual}. We argued that only high-ionization lines ([OIII] 4960, 5008, [NeIII] 3870) show these features while lower-ionization lines ([OII] 3728, H$\beta$) appear to have regular profiles in the same objects. 

In Section \ref{composite} we presented the composite spectrum of type II AGN. We calculated relative line fluxes of 26 emission lines and found that line ratios are very similar to those previously obtained for Seyfert II galaxies. The blue color of the underlying continuum is most likely dominated by young stars in the host galaxies. This conclusion is supported by the previous work (e.g., \citealt{kauf03} and other references in Section \ref{comp_continuum}) that establishes the young stellar population as a general property of host galaxies of powerful AGN.  

In Sections \ref{em_lines} and \ref{comp_em_lines} emission line properties were discussed statistically. We found a strong correlation between the [OII] 3728 and [OIII] 5008 emission lines for all SDSS AGN and for type II AGN in particular. We found evidence that the high-ionization lines ([OIII], [NeIII], [NeV]) are underluminous in type II AGN as compared to type I AGN, for a given luminosity of [OII] 3728, in agreement with some previous studies (e.g., \citealt{naga01} and other references in Sections \ref{em_lines} and \ref{comp_em_lines}). 

In Section \ref{discussion} we presented the correlation between the broad-band magnitudes and the [OIII] 5008 luminosities for all SDSS AGN and suggested how this correlation could be used to estimate the intrinsic luminosities of type II AGN, if the unification model is applied. We estimate that about 50\% of the objects in the sample have intrinsic absolute magnitudes $M_B<-23$ and can be classified as type II quasars based on their intrinsic luminosity. These magnitudes correspond to the luminosity of the [OIII] 5008 emission line of $L$[OIII]$>3\times 10^8 L_{\odot}$. Unlike typical Seyfert II galaxies, whose spectra are dominated by the light from the host galaxy with weak emission lines superimposed on it, the objects from the luminous subsample show very high EWs of emission lines ($50\mbox{\AA}<$EW([OIII])$<1400\mbox{\AA}$). Example spectra of type II quasar candidates are given in Figures \ref{example1}--\ref{example2}.

Several type II AGN with very high [OIII] line luminosities have been described in the literature. The [OIII] luminosity function of low-redshift type II AGN by \citet{hao03} extends up to $L$[OIII]$\simeq 10^9 L_{\odot}$, and the sample by \citet{kauf03} includes four objects with $L$[OIII]$>3\times 10^8 L_{\odot}$. Based on the line luminosity criterion $L$[OIII]$>3\times 10^8 L_{\odot}$, we identify 21 type II quasar candidates among the samples of nearby ($Z\la 0.18$) AGN by \citet{degr92} and \citet{whit92}. We considered all objects in these samples spectroscopically classified by the authors as Seyfert II galaxies and calculated [OIII] 5008 line luminosities in the $\Lambda$-cosmology from the line fluxes and redshifts given in these compilations. The most luminous type II quasar candidates from these samples ($L$[OIII]$>1\times 10^9 L_{\odot}$) are IRAS B094335.8$-$130710, IRAS B105111.2$-$272314, Mrk34, Mrk463 and  Mrk477. 

Estimating the space density of type II AGN or their fraction in the AGN population is a very important task which is, however, beyond the scope of this paper due to the incompleteness of the sample. In our future work, we hope to set useful bounds on the space density of luminous type II AGN based on the small subsample selected by the quasar targeting algorithm. 

We emphasize that multi-wavelength studies of the sample are necessary to support their physical interpretation as obscured AGN. Although technically challenging given the optical flux from the objects in the sample, optical spectropolarimetry would directly confirm the presence of a hidden broad-line region and underlying ionizing continuum through scattered light. Near-IR spectroscopy of the H$\alpha$ region for $Z>0.4$ objects would place the objects on conventional diagnostic diagrams, resolve the questions of underlying broad components, and allow for extinction correction using Balmer decrements. Mid-IR imaging would allow us to compare the spectral energy distribution of our type II AGN candidates with that of known obscured AGN and with theoretical models, while far-IR imaging would allow direct estimates of bolometric luminosities. Column densities of the gas along the line of sight can be inferred from the X-ray spectra. Studies of radio properties would allow comparisons between optically-selected type II AGN and radio galaxies. Such follow-up studies would allow us to estimate the efficiency of optical selection and to study the differences between the populations selected by different techniques. Finally, by studying host galaxies and environments of AGN, one could establish just which properties of the galactic hosts and environments are associated with activity in AGN. Work is underway in all these areas, and we will report on the results in future papers.

The sample of type II AGN presented here was selected based on about a quarter of all spectroscopic data that will have been obtained by the SDSS upon its completion in 2005. We will be able to substantially expand the sample of type II quasar candidates as the SDSS progresses.  

\section*{Acknowledgements}

Funding for the creation and distribution of the SDSS Archive has been provided by the Alfred P. Sloan Foundation, the Participating Institutions, the National Aeronautics and Space Administration, the National Science Foundation, the U.S. Department of Energy, the Japanese Monbukagakusho, and the Max Planck Society. The SDSS Web site is http://www.sdss.org/. 

The SDSS is managed by the Astrophysical Research Consortium (ARC) for the Participating Institutions. The Participating Institutions are The University of Chicago, Fermilab, the Institute for Advanced Study, the Japan Participation Group, The Johns Hopkins University, Los Alamos National Laboratory, the Max-Planck-Institute for Astronomy (MPIA), the Max-Planck-Institute for Astrophysics (MPA), New Mexico State University, University of Pittsburgh, Princeton University, the United States Naval Observatory, and the University of Washington.

NLZ, MAS, MJC, LH and IS acknowledge the support of NSF grant AST00-71091.

The authors are grateful to the referee, Malcolm Smith, for the prompt and encouraging report.

\clearpage

\clearpage
\begin{deluxetable}{ccccccccccccccc}
\pagestyle{empty}
\rotate
\tabletypesize{\tiny}
\tablewidth{0pt}
\setlength{\tabcolsep}{0.03in}
\tablecaption{SDSS type II AGN at $0.3<z<0.83$\label{object_list}}
\tablehead{J2000 & & Redshift, & Target & & & & & & EW, \AA & EW, 
\AA & & & FIRST, & \\
coordinates & DR1 & $Z$ & code & $u^*\pm\sigma_u^*$ & 
$g^*\pm\sigma_g^*$ & $r^*\pm\sigma_r^*$ & $i^*\pm\sigma_i^*$ & 
$z^*\pm\sigma_z^*$ & [OII] 3728 & 
[OIII] 5008 & L[OII] & L[OIII] & mJy & comments}
\startdata 
SDSS J$002531.46-104022.2$ &yes &0.303 &00100 00 &19.94$\pm$0.05 &19.64$\pm$0.01 &19.07$\pm$0.02 &18.98$\pm$0.02 &18.48$\pm$0.04 &40 &59 &8.57 &8.73 &1.41 &\\ 
SDSS J$002711.90+002231.8$ & &0.437 &01000 10 &23.80$\pm$0.85 &21.66$\pm$0.06 &20.17$\pm$0.03 &19.39$\pm$0.02 &18.96$\pm$0.06 &11 & 8 &7.79 &7.92 & &\\ 
SDSS J$002827.78-004218.8$ & &0.418 &00000 10 &22.11$\pm$0.31 &21.34$\pm$0.05 &20.55$\pm$0.04 &19.86$\pm$0.02 &19.56$\pm$0.07 &93 &143 &8.41 &8.75 & &\\ 
SDSS J$002852.87-001433.6$ & &0.310 &00000 10 &23.29$\pm$0.68 &21.46$\pm$0.16 &19.85$\pm$0.08 &19.29$\pm$0.07 &19.01$\pm$0.11 &12 &19 &7.98 &8.43 & &\\ 
SDSS J$004020.31-004033.5$ & &0.568 &00000 10 &21.67$\pm$0.19 &20.86$\pm$0.04 &20.11$\pm$0.03 &19.48$\pm$0.02 &19.05$\pm$0.05 & 7 & 8 &8.03 &8.25 &97.70 &\\ 
SDSS J$004412.87+003606.8$ & &0.502 &00000 10 &22.72$\pm$0.73 &22.05$\pm$0.10 &20.59$\pm$0.04 &19.83$\pm$0.03 &19.39$\pm$0.07 &18 &14 &7.99 &8.27 & &\\ 
SDSS J$005009.81-003900.6$ & &0.729 &00000 10 &21.17$\pm$0.15 &20.68$\pm$0.04 &20.04$\pm$0.03 &19.93$\pm$0.03 &18.91$\pm$0.05 &97 &281 &9.28 &9.94 &4.32 &\\ 
SDSS J$005515.82-004648.6$ & &0.345 &00000 10 &20.86$\pm$0.11 &20.52$\pm$0.03 &19.86$\pm$0.02 &19.54$\pm$0.02 &18.97$\pm$0.05 &17 &35 &7.67 &8.15 & &\\ 
SDSS J$005621.72+003235.8$ & &0.484 &00000 10 &21.98$\pm$0.23 &21.49$\pm$0.06 &20.62$\pm$0.04 &19.54$\pm$0.03 &19.68$\pm$0.09 &70 &303 &8.73 &9.45 &8.60 &asym\\ 
SDSS J$005733.95+001248.3$ &yes &0.377 &00010 10 &23.54$\pm$0.89 &21.33$\pm$0.06 &19.89$\pm$0.03 &19.62$\pm$0.03 &19.06$\pm$0.07 & 9 &13 &7.35 &7.71 & &\\ 
SDSS J$005824.02+005153.3$ & &0.347 &00000 10 &22.59$\pm$0.47 &21.62$\pm$0.06 &20.07$\pm$0.03 &19.37$\pm$0.03 &18.97$\pm$0.05 &14 & 7 &7.65 &7.68 & &\\ 
SDSS J$010255.06-084142.5$ &yes &0.446 &01000 00 &24.42$\pm$1.47 &22.42$\pm$0.12 &20.76$\pm$0.04 &19.87$\pm$0.04 &19.36$\pm$0.09 &16 & 6 &7.83 &7.84 & &\\ 
SDSS J$011228.08-010058.2$ & &0.388 &00000 10 &23.95$\pm$0.74 &22.13$\pm$0.11 &20.61$\pm$0.04 &19.92$\pm$0.04 &19.34$\pm$0.08 &28 &13 &7.73 &7.98 & &\\ 
SDSS J$011429.61+000036.7$ & &0.389 &00000 10 &22.38$\pm$0.25 &21.21$\pm$0.04 &20.06$\pm$0.02 &19.24$\pm$0.02 &18.55$\pm$0.04 &65 &39 &8.50 &8.66 & &\\ 
SDSS J$011522.19+001518.5$ & &0.390 &00000 10 &24.47$\pm$1.18 &21.48$\pm$0.05 &19.79$\pm$0.02 &19.07$\pm$0.01 &18.56$\pm$0.05 &20 &10 &8.02 &8.14 & &\\ 
SDSS J$012032.21-005502.0$ & &0.601 &00000 10 &22.01$\pm$0.19 &21.36$\pm$0.06 &20.79$\pm$0.04 &20.01$\pm$0.04 &20.55$\pm$0.22 &64 &309 &8.44 &9.28 & &\\ 
SDSS J$012325.57+001032.0$ &yes &0.341 &01010 00 &26.11$\pm$0.71 &21.58$\pm$0.07 &19.85$\pm$0.02 &19.32$\pm$0.02 &18.73$\pm$0.05 &15 &14 &7.53 &7.88 & &\\ 
SDSS J$012341.47+004435.9$ & &0.399 &00000 10 &21.65$\pm$0.13 &21.45$\pm$0.05 &20.67$\pm$0.04 &19.69$\pm$0.03 &19.54$\pm$0.08 &119 &379 &8.47 &9.13 &11.83 &\\ 
SDSS J$013401.80+000845.0$ &yes &0.418 &00010 10 &23.59$\pm$0.93 &21.65$\pm$0.07 &20.67$\pm$0.04 &19.83$\pm$0.03 &19.64$\pm$0.10 &94 &171 &8.18 &8.86 & &asym\\ 
SDSS J$013416.34+001413.6$ & &0.555 &00000 10 &22.47$\pm$0.26 &21.53$\pm$0.05 &20.78$\pm$0.04 &19.64$\pm$0.03 &20.04$\pm$0.16 &92 &457 &8.50 &9.53 & &\\ 
SDSS J$013801.57-004946.5$ & &0.433 &00000 10 &21.37$\pm$0.10 &21.01$\pm$0.04 &20.29$\pm$0.03 &19.90$\pm$0.04 &19.32$\pm$0.07 &11 &18 &7.82 &8.29 & &\\ 
SDSS J$013856.14+003437.4$ & &0.478 &00000 10 &21.52$\pm$0.15 &21.21$\pm$0.05 &20.43$\pm$0.04 &19.90$\pm$0.03 &19.69$\pm$0.08 &13 &18 &8.04 &8.29 & &\\ 
SDSS J$013947.31-004305.1$ & &0.443 &00000 10 &22.73$\pm$0.54 &22.47$\pm$0.13 &20.91$\pm$0.05 &20.27$\pm$0.04 &19.78$\pm$0.09 &33 &22 &7.82 &8.00 & &\\ 
SDSS J$014237.49+144117.9$ & &0.389 &00100 00 &20.73$\pm$0.07 &20.01$\pm$0.02 &18.97$\pm$0.01 &18.59$\pm$0.01 &18.26$\pm$0.02 &10 &52 &7.93 &8.76 &n/a &\\ 
SDSS J$014401.31-003220.3$ & &0.599 &00000 10 &21.91$\pm$0.18 &21.47$\pm$0.05 &20.87$\pm$0.04 &19.96$\pm$0.03 &19.51$\pm$0.08 &29 &76 &8.31 &8.91 & &\\ 
SDSS J$014612.80+005112.4$ & &0.622 &00000 10 &23.05$\pm$0.41 &22.71$\pm$0.13 &21.44$\pm$0.07 &20.45$\pm$0.05 &19.88$\pm$0.12 &25 &41 &8.03 &8.51 &1.61 &\\ 
SDSS J$014657.24+005537.2$ & &0.422 &00000 10 &21.24$\pm$0.14 &20.86$\pm$0.04 &20.22$\pm$0.03 &19.66$\pm$0.03 &19.33$\pm$0.07 & 9 &11 &7.77 &8.04 & &\\ 
SDSS J$014828.93-093840.6$ &yes &0.425 &00010 00 &21.69$\pm$0.20 &21.69$\pm$0.08 &20.99$\pm$0.05 &20.49$\pm$0.06 &19.89$\pm$0.15 &12 &24 &7.45 &7.73 & &\\ 
SDSS J$014932.53-004803.7$ & &0.566 &00000 10 &22.55$\pm$0.41 &21.55$\pm$0.06 &20.62$\pm$0.04 &19.66$\pm$0.02 &19.64$\pm$0.08 &54 &228 &8.47 &9.36 &1.03 &\\ 
SDSS J$015716.92-005304.8$ & &0.422 &00000 10 &21.64$\pm$0.12 &21.34$\pm$0.19 &20.55$\pm$0.09 &19.43$\pm$0.07 &19.83$\pm$0.14 &79 &324 &8.75 &9.52 & &\\ 
SDSS J$015813.92-005306.4$ & &0.545 &00000 10 &22.24$\pm$0.20 &21.80$\pm$0.08 &20.87$\pm$0.04 &19.93$\pm$0.04 &19.44$\pm$0.09 &46 &83 &8.29 &8.93 & &\\ 
SDSS J$015911.66+143922.5$ &yes &0.319 &00010 00 &22.55$\pm$0.33 &21.70$\pm$0.06 &20.39$\pm$0.03 &20.40$\pm$0.05 &19.82$\pm$0.10 &38 &301 &7.64 &8.56 &n/a &\\ 
SDSS J$020234.56-093921.9$ &yes &0.302 &01000 00 &21.88$\pm$0.25 &20.53$\pm$0.04 &19.02$\pm$0.02 &18.48$\pm$0.02 &18.09$\pm$0.03 &16 &14 &7.80 &8.12 & &\\ 
SDSS J$021047.01-100152.9$ &yes &0.540 &00100 00 &21.25$\pm$0.12 &20.96$\pm$0.04 &20.43$\pm$0.03 &19.21$\pm$0.02 &20.31$\pm$0.16 &113 &1155 &8.87 &9.79 & &\\ 
SDSS J$021059.66-011145.5$ & &0.384 &00000 10 &21.31$\pm$0.14 &20.81$\pm$0.05 &20.08$\pm$0.02 &19.67$\pm$0.03 &19.09$\pm$0.06 &10 &27 &7.50 &8.10 & &\\ 
SDSS J$021757.82-011324.4$ & &0.375 &00000 10 &22.71$\pm$0.43 &21.49$\pm$0.07 &19.89$\pm$0.02 &19.17$\pm$0.02 &18.65$\pm$0.04 &51 &37 &8.31 &8.55 &1.60 &\\ 
SDSS J$021758.19-001302.7$ &yes &0.344 &00001 00 &21.63$\pm$0.17 &20.52$\pm$0.03 &19.41$\pm$0.01 &19.17$\pm$0.02 &18.67$\pm$0.04 &36 &95 &8.11 &8.75 &1.75 &asym\\ 
SDSS J$021834.42-004610.3$ & &0.372 &00000 10 &21.10$\pm$0.15 &20.55$\pm$0.04 &20.01$\pm$0.03 &19.68$\pm$0.04 &19.15$\pm$0.06 &53 &66 &8.65 &8.85 & &\\ 
SDSS J$021910.76+005919.4$ & &0.346 &00000 01 &21.99$\pm$0.25 &21.23$\pm$0.05 &19.89$\pm$0.04 &19.41$\pm$0.02 &19.09$\pm$0.05 &24 &32 &7.76 &8.09 & &\\ 
SDSS J$021913.61+000926.5$ & &0.350 &00000 01 &21.75$\pm$0.18 &20.74$\pm$0.03 &19.64$\pm$0.02 &19.18$\pm$0.02 &18.79$\pm$0.05 & 8 &10 &7.60 &7.87 & &asym\\ 
SDSS J$022004.64+005908.3$ & &0.413 &00000 01 &22.47$\pm$0.33 &21.49$\pm$0.06 &20.18$\pm$0.03 &19.67$\pm$0.03 &19.37$\pm$0.07 &21 &17 &8.12 &8.26 & &\\ 
SDSS J$022028.74-000641.2$ & &0.364 &00000 01 &22.67$\pm$0.33 &21.65$\pm$0.06 &20.14$\pm$0.03 &19.54$\pm$0.03 &19.05$\pm$0.06 & 7 & 7 &7.15 &7.44 & &asym\\ 
SDSS J$022049.89+001845.8$ & &0.398 &00000 01 &21.96$\pm$0.18 &21.26$\pm$0.04 &20.26$\pm$0.03 &19.83$\pm$0.03 &19.42$\pm$0.10 &10 &17 &7.72 &8.04 & &\\ 
SDSS J$022146.36-004511.9$ & &0.357 &00000 01 &21.40$\pm$0.15 &21.00$\pm$0.04 &20.03$\pm$0.02 &19.65$\pm$0.02 &19.15$\pm$0.06 &13 &25 &7.57 &7.95 & &\\ 
\enddata

\tablecomments{DR1 column indicates whether the object is in the SDSS Data Release 1 \citep{abaz03}. Redshifts were calculated based on the [OII] 3728 emission line. The first five digits of the Target code show whether the object was targeted on one of the main plates by Galaxy, LRG, Quasar, High-redshift quasar and Serendipity algorithms, in that order. A digit of 0 means the object was not targeted by this method, while 1 means targeted. The last two digits indicate whether the object was targeted on one of the DSES plates or on one of the Special plates, as described in Section \ref{data}. We use PSF magnitudes to minimize contribution from the host galaxies. Equivalent widths are calculated in the rest frame, relative to the total continuum. Line luminosities are calculated based on the best-fit Gaussian fitting, assuming a $h=0.7$, $\Omega_m=0.3$, $\Omega_{\Lambda}=0.7$ cosmology, and are given as $\log(L_{\rm line}/L{\odot})$. In the column `FIRST' we list the integrated fluxes at 20 cm in mJy, if the object is matched within 3\arcsec; there is no entry if the object was not detected ($F_{\nu}$(20cm)$<$1mJy). If the field of the object was not observed by the FIRST survey, `n/a' is listed. In the column ``comments'', the flag `asym' is set for objects showing asymmetric line profiles, see Section \ref{unusual}. The complete version of this table is in the electronic edition of the Journal.}
\end{deluxetable}

\clearpage
\begin{deluxetable}{cccccc}
\tabletypesize{\footnotesize}
\tablecaption{Relative emission line fluxes of the type II AGN composite spectrum\label{relative}}
\tablehead{&vacuum &air &EWs, & rel. flux, & rel. flux, \\
Line ID &wavelength, \AA &wavelength, \AA & \AA & 100$\times F/F$([OII]) & type I QSOs }
\startdata
MgII	 &2800.0	 &2799.2	 &5.2$\pm$0.8	 &27.7$\pm$ 0.7	 &\\ 
OIII	 &3133.7	 &3132.8	 &1.1$\pm$0.1	 &4.6$\pm$ 0.1	 &\\ 
HeII	 &3204.0	 &3203.1	 &0.8$\pm$0.1	 &3.5$\pm$ 0.1	 &\\ 
$[$NeV$]$	 &3346.8	 &3345.8	 &1.4$\pm$0.3	 &5.8$\pm$ 1.1	 &27.9\\ 
$[$NeV$]$	 &3426.8	 &3425.9	 &5.6$\pm$0.1	 &22.2$\pm$ 0.2	 &95.9\\ 
$[$OII$]$	 &3728.4	 &3727.3	 &24.5$\pm$0.3	 &100.0$\pm$ 0.2	 &100.0\\ 
$[$NeIII$]$	 &3870.2	 &3869.1	 &7.6$\pm$0.2	 &34.8$\pm$ 0.4	 &81.8\\ 
HeI+H8	 &3889.0	 &3887.9	 &0.2$\pm$0.2	 &1.0$\pm$ 0.9	 &\\ 
$[$NeIII$]$+H$\varepsilon$	 &3968.9	 &3967.8	 &1.2$\pm$0.3	 &5.9$\pm$ 1.5	 &\\ 
$[$FeV$]$+$[$SII$]$	 &4073.6	 &4072.5	 &1.1$\pm$0.1	 &5.9$\pm$ 0.1	 &9.2\\ 
H$\delta$	 &4102.9	 &4101.7	 &0.6$\pm$0.1	 &3.1$\pm$ 0.3	 &\\ 
H$\gamma$	 &4341.7	 &4340.5	 &2.4$\pm$0.1	 &13.1$\pm$ 0.1	 &\\ 
$[$OIII$]$	 &4364.4	 &4363.2	 &1.6$\pm$0.1	 &8.7$\pm$ 0.4	 &26.0\\ 
HeII	 &4687.0	 &4685.7	 &1.3$\pm$0.1	 &8.1$\pm$ 0.5	 &\\ 
H$\beta$	 &4862.7	 &4861.4	 &5.4$\pm$0.1	 &34.1$\pm$ 0.3	 &\\ 
$[$OIII$]$	 &4960.3	 &4958.9	 &12.1$\pm$0.2	 &78.0$\pm$ 2.0	 &162.4\\ 
$[$OIII$]$	 &5008.2	 &5006.8	 &29.4$\pm$1.9	 &187.3$\pm$ 14.0	 &588.2\\ 
$[$NI$]$	 &5200.5	 &5199.1	 &0.9$\pm$0.1	 &5.3$\pm$ 0.4	 &23.2\\ 
HeI	 &5877.3	 &5875.7	 &0.8$\pm$0.1	 &4.8$\pm$ 0.1	 &\\ 
$[$OI$]$	 &6302.0	 &6300.3	 &3.2$\pm$0.1	 &19.6$\pm$ 0.6	 &42.3\\ 
$[$OI$]$	 &6365.5	 &6363.8	 &1.1$\pm$0.1	 &6.9$\pm$ 0.4	 &51.2\\ 
$[$NII$]$	 &6549.9	 &6548.0	 &8.1$\pm$0.3	 &50.1$\pm$ 1.2	 &46.1\\ 
H$\alpha$	 &6564.6	 &6562.8	 &22.9$\pm$0.8	 &141.2$\pm$ 6.6	 &\\ 
$[$NII$]$	 &6585.3	 &6583.4	 &24.3$\pm$0.9	 &150.2$\pm$ 4.0	 &196.5\\ 
$[$SII$]$	 &6718.3	 &6716.4	 &6.6$\pm$0.4	 &40.7$\pm$ 2.4	 &65.3\\ 
$[$SII$]$	 &6732.7	 &6730.8	 &5.7$\pm$0.4	 &35.1$\pm$ 2.2	 &57.6\\ 
\enddata

\tablecomments{Line identifications were taken from numerous sources, see \citet{vand01}. Vacuum wavelengths were taken from the Atomic Line List; for multiplets, centroid wavelengths were taken from \citet{vand01}. Air wavelengths were calculated from vacuum wavelengths based on IAU standards \citep{mort91}. We subtract the host galaxy to reveal weak emission features superimposed on absorption features from the host galaxy and fit the lines with Gaussian profiles to obtain relative fluxes. EWs are calculated relative to the total (unsubtracted) continuum. Errors include errors of fitting the lines with Gaussians as well as the differences in line parameters as measured from the [OII] 3728 and [OIII] 5008-based composite spectra. Relative fluxes of forbidden emission lines of broad-line quasars from \citet{vand01} are given in the last column for comparison.}
\end{deluxetable}

\end{document}